\documentclass[twocolumn,letterpaper,amsmath,amssymb,floatfix,aps,superscriptaddress]{revtex4}

\def\ul#1#2{\textstyle{\frac{#1}{#2}}}
\def\tr{{\mathrm{Tr}}}
\newcommand {\vct}[1] {\mathbf {#1}}

\usepackage{graphicx}
\usepackage{dcolumn}  
\usepackage{bm}  
\usepackage{epsfig}

\begin{document}
\setlength\arraycolsep{2pt}

\title{Non-monotoic fluctuation-induced interactions between dielectric slabs carrying charge disorder}

\author{Jalal Sarabadani}
\affiliation{Department of Physics, University of Isfahan, Isfahan
81746, Iran}
\affiliation{Department of Theoretical Physics, J. Stefan
Institute, SI-1000 Ljubljana, Slovenia} 

\author{Ali Naji}
\affiliation{Department of Applied Mathematics and Theoretical Physics, Centre for Mathematical Sciences, University of Cambridge, Cambridge CB3
0WA, United Kingdom}

\author{David S. Dean}
\affiliation{Laboratoire de
Physique Th\'eorique (IRSAMC),Universit\'e de Toulouse, UPS and CNRS,  F-31062 Toulouse, France}

\author{Ron R. Horgan}
\affiliation{Department of Applied Mathematics and Theoretical Physics, Centre for Mathematical Sciences, University of Cambridge, Cambridge CB3
0WA, United Kingdom}

\author{Rudolf Podgornik}
\affiliation{Department of Theoretical Physics, J. Stefan
Institute, SI-1000 Ljubljana, Slovenia} 
\affiliation{Institute of
Biophysics, School of Medicine and Department of Physics, Faculty
of Mathematics and Physics, University of Ljubljana, SI-1000
Ljubljana, Slovenia}

\begin{abstract}
We investigate the effect of monopolar charge disorder on the classical   
fluctuation-induced interactions between randomly charged net-neutral dielectric slabs and
discuss various generalizations of recent results (A. Naji {\em et al.}, Phys. Rev. Lett. {\bf 104}, 060601 (2010)) 
to highly inhomogeneous dielectric systems with and without statistical disorder correlations. 
We shall focus on the specific case of two generally dissimilar plane-parallel slabs, which interact 
across vacuum or an arbitrary intervening dielectric medium. 
Monopolar charge disorder is considered to be present on the bounding surfaces and/or 
in the bulk of the slabs, may be in general quenched or annealed and may possess a finite 
lateral correlation length reflecting possible `patchiness' of the random charge distribution. 
In the case of quenched disorder, the bulk disorder is shown to give rise to an additive long-range 
contribution to the total force, which decays as the  inverse distance between the slabs and may be attractive or repulsive depending 
on the dielectric constants of the slabs. By contrast, the force induced by annealed disorder  in general combines with 
the  underlying van der Waals forces in a non-additive fashion and the net force decays as an inverse cube law at large separations. 
We show however that in the case of two dissimilar slabs 
the net effect due to the interplay between the disorder-induced  
and the pure van der Waals interactions  can lead to a variety of unusual non-monotonic interaction profiles between the dielectric slabs.
In particular,  when the intervening medium has a larger 
dielectric constant than the two slabs, we find that the net interaction can become repulsive and exhibit a potential barrier, while the 
underlying van der Waals force is attractive. On the contrary, 
when the intervening medium has
a dielectric constant in  between that of the two slabs, 
the net interaction can become attractive and exhibit a free energy  
minimum, while the  pure 
van der Waals force is repulsive. Therefore, the charge disorder, if present, can drastically alter the effective interaction between net-neutral objects.
\end{abstract}
\maketitle

\section{Introduction}

One of the most persistently made assumptions in the standard theory of (bio)colloid stability is
that of a uniform  charge distribution on macromolecular surfaces \cite{DLVO}. There are 
nevertheless many instances where charge patterns on macromolecular surfaces are inhomogeneous,  
exhibiting  even a highly disordered spatial distribution. Among most notable examples are
surfactant-coated surfaces \cite{Klein} and random polyelectrolytes and polyampholytes \cite{rand_polyelec}. 
The macromolecular charge pattern in these systems is often frozen, or quenched, meaning that 
the charge distribution does not evolve after the assembly or fabrication of the material. This is the case
if the interaction between the  surface charge carriers and the surfaces is of a chemical nature involving 
interactions that far exceed the thermal energy scale $k_{\mathrm{B}}T$. Alternatively charge distributions can 
exhibit various degrees of annealing when interacting with other macromolecules in aqueous solutions as is, 
for example, the case for charge regulation of contact surfaces bearing weak 
acidic groups in aqueous solutions \cite{ParsegianChan}. 

Recently it has also been realized that 
ultrahigh sensitivity experiments on Casimir (zero temperature and ideally polarizable surfaces) and 
van der Waals (finite temperature and non-ideally polarizable surfaces) interactions between surfaces 
{\em in vacuo} \cite{bordag,kim} can be properly understood only if one takes into account the disordered 
nature of charges on and within the interacting surfaces  \cite{kim,speake, PRL}.
Possible causes of  sample- and history-specific charge disorder in this case include 
the  patch effect, where the variation  of the local crystallographic axes of the exposed surface 
of a  clean polycrystalline sample can lead to a variation of the local surface potential \cite{barrett}. 
Amorphous films deposited on crystalline substrates can also show a similar type of surface charge 
disorder, showing a grain structure of dimensions sometimes larger than the thickness of the 
deposited surface film \cite{liu}. On the other hand, adsorption of various contaminants  can influence the nature and type of the surface charge disorder.

Since the nature and distribution of the charge disorder in any of the force experiments is in general seldom known, we pursue a strategy of  assessing  the consequences of different {\em a priori} models of 
the distribution of charge disorder. In what follows we will thus assume that the charge
disorder stems from randomly distributed {\em monopolar charges} which may be present both 
in the bulk and/or on the interacting surfaces and can be either annealed or quenched.  In the  quenched case, the disorder charges
are  frozen, whereas in the annealed case,
the disorder charges are subject to thermal fluctuations at ambient 
temperature and can thus adapt themselves in order to minimize 
the free energy of the system. We 
do not deal with effects due to  disorder in the dielectric response of the interacting 
media \cite{randomdean}, which presents an additional source of disorder meriting further study. 
We have shown in our previous works \cite{ali-rudi,mama,PRL} that 
the type and the nature of the charge disorder induces marked changes in the properties of the total 
interaction between apposed bodies and the force arising from the presence of disordered charges can dominate the underlying Casimir--van der Waals (vdW)  effect at sufficiently large separations. The analysis of the {\em interaction fingerprint} of the charge disorder can be useful in assessing whether the experimentally observed interactions can be interpreted  in terms of disorder effects or are due to pure Casimir--vdW interactions.  
In this study we will continue with the assessment of this interaction fingerprint 
in the case two
semi-infinite net-neutral dielectric slabs (separated by a layer of vacuum or an arbitrary unstructured dielectric material as shown in Fig. \ref{fig:schematic}), 
by generalizing our
formalism to include two different mechanisms:  
\begin{enumerate}
\item[i)] the  ``patchy" distribution of the charge disorder, which may arise due to
finite correlations between the carriers of the  charge disorder.  This is a rather realistic assumption in view of the  graininess observed in interacting 
surfaces \cite{barrett, liu}. 
\item[ii)] the dielectric inhomogeneity effects in highly asymmetric systems, where the two
slabs and the intervening medium may in general have different dielectric properties.
\end{enumerate}


For interacting bodies carrying 
no {\em net} charge, the part of the interaction due to monopolar  {\em quenched charge} disorder can be seen from a full quantum field-theoretical formalism 
\cite{unpublished} to be coupled directly to the zero-frequency or {\em classical} vdW interaction (corresponding to the zero-frequency Matsubara modes of the electromagnetic field). The situation would be more complicated in the case 
of annealed disorder if one is to account for the higher-order Matsubara frequencies  
and a general treatment is missing at present. In this work, we shall 
examine the effects of quenched and annealed 
charge disorder by focusing specifically on 
the zero-frequency vdW interaction. It is well known that 
the higher frequency Matsubara terms in the total vdW interaction free energy become relatively unimportant at sufficiently high temperatures and/or sufficiently large intersurface separations \cite{bordag}. 
In this paper we will restrict ourselves to this regime which is also most relevant 
experimentally \cite{kim} and consider 
the short-distance quantum effects  elsewhere \cite{unpublished}.

As we shall demonstrate, 
the interplay and competition of the underlying
vdW effect between the two semi-infinite slabs and the interaction induced by
the charge disorder can give rise to a variety of different 
interaction profiles, including,  most notably, {\em non-monotonic} interactions
as a function of the inter-slab distance. 
 In particular, when the two dielectrics
interact across a medium of higher dielectric constant, the
disorder force can become strongly repulsive and thus generate a
potential barrier when combined with the vdW force, which is attractive in that case. While, in the situation where the intervening medium  has a
dielectric constant in between that of the two slabs  \cite{vdw_repulsive}, the 
disorder can generate an attractive force, which
can balance the repulsive vdW force, leading thus to a stable bound state
between the two dielectric slabs. 
Therefore, the charge disorder, if present, can provide an intrinsic
mechanism to stabilize interactions between net-neutral bodies.
These features emerge because the
disorder-induced interactions typically decay more weakly with
the separation than the vdW interaction and depend strongly on the
dielectric inhomogeneities in the system.

The above features vary depending on whether the disorder is assumed to be
quenched or annealed. 
The effects  due to annealed charge disorder can be distinctly 
different from those of the quenched charge disorder.  
In the case of annealed charges, the effective interactions
turn out to decay more rapidly and in fact in a similar way 
as the pure vdW interaction, whereas in the case of 
quenched charges the net interactions  decay more 
weakly with the separation, i.e., as $\sim D^{-\alpha}$ where 
$\alpha=1$ if the disorder is present  in the bulk of the slabs 
and $\alpha=2$ if the disorder is confined to the bounding surfaces
of the slabs.

The organization of the paper is as follows:
In Section \ref{sec:model}, we introduce the details of our model and
the formalism used in our study. The general results are derived
in Sections \ref{sec:quenched} and \ref{sec:annealed}  for the
case of quenched and annealed disorder, respectively.
We shall proceed with the analysis of the effective interaction between
two identical slabs in Section \ref{sec:symmetric}, where we
shall focus mainly on the effects due to spatial correlation in the distribution
of the charge disorder. 
In Section \ref{sec:asymmetric}, we study the interaction in dielectrically 
asymmetric
systems by considering two dissimilar dielectric slabs that interact
across an arbitrary dielectric medium. The results and limitations
of our study are summarized in Section \ref{sec:discussion}.

\section{Model}
\label{sec:model}

\begin{figure}[t]
\includegraphics[angle=0,width=8.5cm]{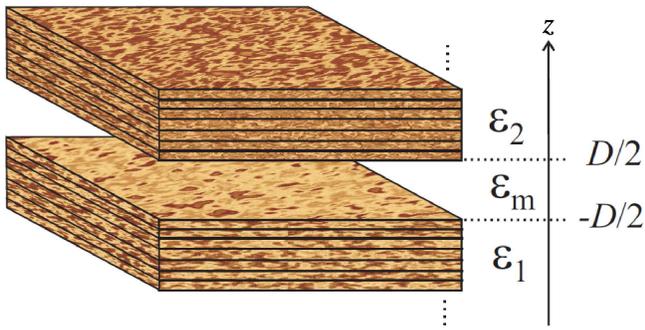}
\caption{ (Color
online) We consider two semi-infinite net-neutral slabs
(half-spaces) of dielectric constant $\varepsilon_1$ and
$\varepsilon_2$ interacting across a medium of dielectric constant
$\varepsilon_m$. The monopolar charge disorder  (shown
schematically by small light and dark patches) is distributed as
random patches  of finite typical size (correlation length) in a
layered structure in the bulk of the slabs and on the two bounding
surfaces at $z = \pm D/2$. It may be either quenched or annealed.}
\label{fig:schematic}
\end{figure}

Let us consider two semi-infinite slabs of dielectric constants
$\varepsilon_{1}$ and $ \varepsilon_{2}$ and temperature $T$ with
parallel planar inner surfaces (of infinite area $S$) located
normal to the $z$ axis at $z=\pm D/2$ (see Fig.
\ref{fig:schematic}). The inner gap is filled with a material of
dielectric constant $ \varepsilon_{m}$. The inhomogeneous
dielectric constant profile for this system is thus given by
\begin{eqnarray}
\varepsilon({\mathbf  r})  &=&  \left\{
\begin{array}{ll}
     \varepsilon_1&   \quad z<-D/2,\\
         \varepsilon_m   & \quad   |z|<D/2,\\
     \varepsilon_2  &   \quad  z>D/2.
\end{array}
 \label{eq:epsilon}
\right.
\end{eqnarray}
We shall assume that the two dielectric slabs have a disordered
{\em monopolar charge} distribution, $\rho(\vct r)$, which may
arise from randomly distributed charges residing on the bounding
surfaces [$\rho_s(\vct r)$] and/or in the bulk [$\rho_b(\vct r)$],
i.e., $\rho(\vct r)=\rho_s(\vct r)+\rho_b(\vct r)$. The charge
disorder is assumed to be distributed according to a Gaussian
weight with zero mean (i.e., the slabs are {\em net neutral}), and
the two-point correlation function
\begin{equation}
\langle \! \langle  \rho(\vct r) \rho(\vct r ') \rangle   \!
\rangle = {\mathcal G}({\boldsymbol \varrho} - {\boldsymbol \varrho}'; z) \delta (z- z'),
\label{charge_correlation}
\end{equation}
where  $\langle \! \langle  \cdots  \rangle   \! \rangle$ denotes
the average over all realizations of the charge disorder
distribution, $\rho(\vct r)$. We have thus assumed that there are
no spatial correlations in the perpendicular direction, $z$,
while, in the lateral directions ${\boldsymbol \varrho} = (x, y)$
(in the plane of the dielectrics), we have a finite statistically
invariant correlation function whose specific form may depend on
$z$ as well. This implies that the charge disorder is distributed
in general as random ``patches" in a layered structure in the bulk
of the slabs as well as on the bounding surfaces.

The total correlation function can be written as the sum of the
surface ($s$) and bulk ($b$) contributions
\begin{equation}
{\mathcal G}({\boldsymbol \varrho} - {\boldsymbol \varrho}'; z)  = g_s(z)c_{s} ({\boldsymbol \varrho} - {\boldsymbol \varrho}'; z)+
 g_b(z)c_{b} ({\boldsymbol \varrho} - {\boldsymbol \varrho}'; z).
\label{eq:bs_decomp}
 \end{equation}
 For the slab geometry, we generally assume that the lateral correlation
 functions may be different for the two slabs, i.e.
\begin{eqnarray}
c_{s}({\mathbf x}; z) &=& \left\{
\begin{array}{ll}
     c_{1s}({\mathbf x})  & \quad  z=-D/2,\\
     c_{2s}({\mathbf x})  & \quad  z=D/2,
\end{array}
\right.
 \label{eq:c_s}
 \\
c_{b}({\mathbf x}; z) &=&  \left\{
\begin{array}{ll}
     c_{1b}({\mathbf x})  & \quad  z<-D/2,\\
          0      & \quad  |z|<D/2,\\
     c_{2b}({\mathbf x})  & \quad  z>D/2,
\end{array}
\right.
\label{eq:c_b}
\end{eqnarray}
and that the surface and bulk variances are given by
\begin{eqnarray}
 g_s(z) &=& e_0^2 [g_{1s}\delta(z + D/2) +  g_{2s}\delta(z - D/2)],
 \label{eq:g_s}
 \\
g_b(z) &=&  \left\{
\begin{array}{ll}
     g_{1b}e_0^2  & \quad  z<-D/2,\\
          0      & \quad  |z|<D/2,\\
     g_{2b}e_0^2  & \quad  z>D/2.
\end{array}
 \label{eq:g_b}
\right.
\end{eqnarray}

The lateral correlation between two given points is typically
expected to decay with their separation over a finite {\em
correlation length} (``patch size"), which could in general be
highly material or sample specific. However, the main aspects of
the patchy structure of the disorder can be investigated by
assuming
simple generic models with, for instance, a Gaussian  
or an exponentially decaying correlation function.
Without loss of generality, we shall choose an exponentially decaying correlation function according to
 the two-dimensional Yukawa form
 \begin{equation}
 c_{i\alpha}({\mathbf x}) = \frac{1}{2 \pi \xi_{i\alpha}^2}\,\mathrm{K}_0\!\left(\frac{|{\mathbf x}|}{\xi_{i\alpha}}\right),
\label{eq:Lorentzian}
 \end{equation}
where $\xi_{i\alpha}$ represents the correlation length for the
bulk or surface disorder ($\alpha=b, s$) in the $i$-th slab ($i=1,
2$). The case of a completely {\em uncorrelated} disorder \cite{PRL}
follows as a special case for $\xi_{i\alpha} \rightarrow 0$ from
our formalism. We should emphasize that the correlation assumed for the
bulk disorder is only present in the plane of the slab surfaces and 
not in the direction perpendicular to them. This assumption is wholly
justified only for {\em layered materials}. In all the other cases of the bulk disorder one
would normally expect the same correlation length in the direction perpendicular 
to the surfaces. We will deal with this model in a separate publication.

\section{Formalism}
\label{sec:formalism}

The partition function for the classical vdW interaction (the zero-frequency Matsubara
modes of the electromagnetic field)  may be written as a functional integral over the  scalar field $\phi(\vct r)$,
\begin{equation}
{\mathcal Z}[\rho(\vct r)] = \int [{\mathcal D}\phi(\vct r)] \,\,e^{- \beta {\mathcal S}[\phi(\vct r); \rho(\vct r)]},
\label{eq:Z}
\end{equation}
with $ \beta = 1/k_{\mathrm{B}}T$ and  the effective action
\begin{equation}
{\mathcal S}[\phi(\vct r); \rho(\vct r)] =  \int {\mathrm{d}} \vct r\, \big[{\ul{1}{2}}  \varepsilon_{0}
\varepsilon({\mathbf  r})\, (\nabla \phi(\vct r))^{2}   + {\mathrm{i}} \, \rho(\vct r) \phi(\vct r)\big].
\label{act-1}
\end{equation}
The above partition function can be used to evaluate averaged quantities such as
the effective interaction between the dielectric bodies. However, since the charge distribution, $ \rho(\vct r)$,
is disordered, it is necessary to average the partition function over different realizations
of the charge distribution. The averaging procedure differs depending on the nature of the disorder.
In what follows, we consider two idealized cases of either completely quenched or completely annealed 
disorder \cite{dotsenko1} (the intermediate case of partially annealed disorder is also analytically tractable 
\cite{mama} but will not be considered here).

\subsection{Correlated quenched disorder}
\label{sec:quenched}

The quenched disorder corresponds to the situation where the disorder charges are frozen and can not fluctuate and
equilibrate with other degrees of freedom in the system. As it is well known,  the disorder 
average in this case must be taken over the sample
free energy, $\ln {\mathcal Z}[\rho(\vct r)]$, in order to evaluate the averaged quantities \cite{dotsenko1}. 
Therefore, the free energy of the quenched model is given by
\begin{equation}
\label{eq:quenched_freen}
 \beta {\mathcal F}_{\mathrm{quenched}}
    = -  \langle \!   \langle  \ln {\mathcal Z}[\rho(\vct r)]  \rangle   \! \rangle.
\end{equation}
The Gaussian integral in Eq. (\ref{eq:Z}) as well as the disorder
average can be evaluated straightforwardly in this case yielding 
\begin{eqnarray}
\label{eq:F_quenched}
\beta{\mathcal F}_{\mathrm{quenched}}  &=& {\ul{1}{2}}\tr \ln
G^{-1}({\vct r}, {\vct r}')\\
&+& {\ul{\beta}{2}}  \int\! {\mathrm{d}}{\mathbf r}\,{\mathrm{d}}{\mathbf r}'\, {\mathcal G}({\boldsymbol \varrho} - {\boldsymbol \varrho}'; z)
\delta(z-z') G({\vct r}, {\vct r}'), \nonumber
\end{eqnarray}
where $G({\vct r}, {\vct r}')$ is the Green's function defined via
\begin{equation}
 \varepsilon_0 \nabla\cdot [\varepsilon(\vct r) \nabla  G(\vct r, \vct r')] = -\delta(\vct r - \vct r').
 \label{green1}
\end{equation}
Note that the free energy (\ref{eq:F_quenched}) of the quenched model 
is expressed in terms
of different additive contributions, i.e.
\begin{equation}
{\mathcal F}_{\mathrm{quenched}}= {\mathcal F}_{\mathrm{vdW}} + {\mathcal F}_{b} + {\mathcal F}_{s}. 
 \label{free_energy_quenched}
\end{equation}
The first term above is nothing but the usual contribution from the zero-frequency vdW interaction, 
\begin{equation}
\beta {\mathcal F}_{\mathrm{vdW}} =  {\ul{1}{2}}\tr \ln G^{-1}({\vct r}, {\vct r}'),
\end{equation}
which is always present between neutral
dielectrics even in the absence of any monopolar charge disorder. The second and the third terms
represent contributions from the  bulk and surface disorder ($\alpha=b, s$)
\begin{equation}
 \beta {\mathcal F}_{\alpha} =  {\ul{\beta}{2}} \int\! {\mathrm{d}}{\mathbf r}\,{\mathrm{d}}{\mathbf r}'\,g_{\alpha}(z) c_{\alpha}({\boldsymbol \varrho}  -
{\boldsymbol \varrho}'; z) \delta(z-z')G({\vct r}, {\vct r}').
\label{eq:Quenched_each}
\end{equation}

The quenched expression  (\ref{eq:F_quenched}) is valid for any
arbitrary disorder correlation function $ {\mathcal G}({\boldsymbol \varrho} - {\boldsymbol \varrho}'; z) $ and dielectric constant profile $\varepsilon({\vct r})$.
In what follows, we shall focus on the particular case of planar dielectrics by employing Eqs. (\ref{eq:epsilon}), (\ref{eq:bs_decomp})-(\ref{eq:g_b}).
In this case,  the zero-frequency vdW contribution is obtained (per $k_{\mathrm{B}}T$ and unit area) as \cite{bordag}
\begin{equation}
\frac{\beta {\mathcal F}_{\mathrm{vdW}}}{S} =  {\frac{1}{2}}\!
\int\!\!\frac{{\mathrm{d}}^2Q}{(2\pi)^2} \ln{(1 - \Delta_1
\Delta_2 \,
e^{- 2 Q D})}. 
\label{eq:F_vdW}
\end{equation}
The force associated with this contribution, $f_{\text{vdW}} = - \partial
{\mathcal F}_{\mathrm{\text{vdW}}} / \partial D$, follows as
\begin{equation}
\frac{\beta f_{{\text {vdW}}}}{S} =
-\frac{{\mathrm{Li}}_3(\Delta_1 \Delta_2)}{8\pi D^3}.
\label{eq:vdW_full}
\end{equation}
The dielectric jump parameters are  defined as
\begin{equation}
\Delta_i = \frac{\varepsilon_{i}  - \varepsilon_{m}}{\varepsilon_{i}  +
\varepsilon_{m}}
\end{equation}
at each of the bounding surfaces ($i=1, 2$), and ${\mathrm{Li}}_3(\cdot)$ is the
trilogarithm function defined by
\begin{equation}
{\rm Li}_3(z) = \sum_{n=1}^\infty {z^n\over n^3}.
\end{equation}
The dielectric constants of the system enter the expression (\ref{eq:F_vdW}) via the terms $\Delta_1$ and $\Delta_2$. If $\Delta_1 \Delta_2 >0$ then the vdW interaction between the slabs is
attractive, otherwise the vdW  force is repulsive.

The expression in Eq. (\ref{eq:Quenched_each}) can be calculated in a straightforward manner.
We obtain the bulk  disorder contribution as
\begin{eqnarray}
\label{eq:free_energy_g_b_quenched}
\frac{ \beta {\mathcal F}_{b}}{S}&& =   - \ell_{\mathrm{B}} \varepsilon_m
\times  \\
&&\hspace{-1.1cm} \int  \! \frac{{\mathrm{d}}Q}{Q}
\frac{e^{-2 Q D}}{1 - \Delta_1 \Delta_2 e^{-2 Q D}} \bigg[ \!
\frac{g_{1b} c_{1b}(Q)}{ (\varepsilon_1 \! + \! \varepsilon_m)^2 }
\Delta_2 \! + \! \frac{g_{2b} c_{2b}(Q)}{ (\varepsilon_2 \! + \!
\varepsilon_m)^2 }
\Delta_1 \! \bigg]  , \nonumber
\end{eqnarray}
and the surface disorder contribution as
\begin{eqnarray}
\label{eq:free_energy_g_s_quenched}
\frac{ \beta {\mathcal F}_{s}}{S}&& =   - 2 \ell_{\mathrm{B}} \varepsilon_m
\times \\
&&\hspace{-1.1cm} \int  \! \mathrm{d} Q \frac{e^{-2 Q
D}}{1 - \Delta_1 \Delta_2 e^{-2 Q D}} \bigg[ \! \frac{g_{1s}
c_{1s}(Q)}{ (\varepsilon_1 \! + \! \varepsilon_m)^2 } \Delta_2
\!+\! \frac{g_{2s} c_{2s}(Q)}{ (\varepsilon_2 \! + \!
\varepsilon_m)^2 }
\Delta_1 \! \bigg]  , \nonumber
\end{eqnarray}
at all separations $D$, where 
\begin{equation}
\ell_{\mathrm{B}}= \beta e_0^2 /(4 \pi
{\varepsilon_0}) 
\end{equation}
 is the Bjerrum length in vacuum  ($\ell_{\mathrm{B}}\simeq 56.8 $~nm at room temperature),
and $c_{i\alpha} (Q)$ is the Fourier transform of the correlation function $c_{i\alpha} ( {\mathbf x})$. For the particular
form of $c_{i\alpha} ( {\mathbf x})$ chosen in Eq. (\ref{eq:Lorentzian}), we have a Lorentzian Fourier
transform  (for $i=1, 2$ and $\alpha=b, s$) as
\begin{equation}
c_{i\alpha}(Q) = \frac{1}{\xi_{i\alpha}^2 Q^2 + 1}.
 \label{eq:Fourier_correlation_function}
\end{equation}

The total force between the dielectric slabs carrying quenched charge disorder
thus follows from
$f_{\mathrm{quenched}} = -\partial {\mathcal F}_{\mathrm{quenched}}/\partial D$ and the preceding free energy
expressions as
\begin{eqnarray}
\label{eq:force_quenched}
\frac{ \beta {f}_{\mathrm{quenched}}}{S} &=& -\frac{{\mathrm{Li}}_3(\Delta_1 \Delta_2)}{8\pi D^3} -2 \ell_{\mathrm{B}}
\varepsilon_m  \times \\
&& \hspace{-2.2cm}  \!\! \int_{0}^{\infty} \!\!\!\!\!
\frac{{\mathrm{d}}Q~ e^{-2 Q D}}{\big(1 \! - \! \Delta_1 \Delta_2
e^{-2 Q D}\big)^2} \bigg[ \! \frac{g_{1b} c_{1b}(Q)}{
(\varepsilon_1 \! + \! \varepsilon_m)^2 } \Delta_2 \!+\!
\frac{g_{2b} c_{2b}(Q)}{
(\varepsilon_2 \! + \! \varepsilon_m)^2 } \Delta_1 \! + \nonumber\\
&& \hspace{+0.2cm} + \frac{2 Q g_{1s} c_{1s}(Q)}{ (\varepsilon_1
\!+\! \varepsilon_m)^2 } \Delta_2 \!+\! \frac{2 Q g_{2s}
c_{2s}(Q)}{ (\varepsilon_2 \!+\! \varepsilon_m)^2
} \Delta_1 \bigg]. \nonumber
\end{eqnarray}

We notice that the quenched disorder contribution to the force has a number of
interesting features. We see that the forces generated due to the charge disorder in slab 
1 and slab 2 are independent and additive. If we remove the charge disorder from either
slab we see there is still a contribution coming from the other slab. This is because the 
charge distributions on average do not interact with each other. As the charge distribution in
opposing slabs is uncorrelated, the average force on a charge in slab 1 due to a charge in slab 2
is zero as the charge in slab two is equally likely to have a positive charge as a negative charge.
In fact the charges in slab 1 are only correlated with their {\em image charges} in slab 2, this means that
on average the charge distribution in slab 1 only interacts with its image in slab 2 and vice versa.    
The {\em sign} of the interaction between charges in slab 1 and their images in slab 2 depends on
$\Delta_2$. If $\Delta_2$ is positive, that is the dielectric constant of slab 2 is greater than that of the intervening material, $\varepsilon_2> \varepsilon_m$, then the force due to the charge in slab 1
is attractive. Therefore, the contribution of the charge in each slab to the net force depends on 
the dielectric contrast between the opposing slab and the intervening medium. Thus, while
the vdW contribution depends on the product $\Delta_1\Delta_2$, the contribution
from the quenched charge distribution in slab 1 depends on variance $g_{1s,b}$ of the charge on the  surface  or 
bulk of this slab but on $\Delta_2$ of the slab 2. This leads to a rich phenomenology of possible net interactions. For example, if  $\Delta_1$ and $\Delta_2$ are both positive, the vdW
force is attractive and so are the forces due to both charge distributions. However, if $\Delta_1$ and
$\Delta_2$ are negative the vdW force is again attractive but the force due to both  quenched
charge distributions is repulsive. 
Also, note that while the vdW term shows a standard $1/D^3$ decay with the surface separation $D$, the disorder contribution turns out to have a much weaker decay with the separation \cite{PRL} as we shall analyze further 
in the forthcoming sections.
Therefore, the interplay between these different contributions can lead to characteristically different interaction profiles depending on the system parameters. 

\subsection{Correlated annealed disorder}
\label{sec:annealed}

In the annealed model, disorder charges are assumed to fluctuate in thermal equilibrium with the
rest of the system, and thus in particular the charge distribution in
the two slabs can adapt itself to minimize the free energy
of the system. In this case, the disorder  degrees of freedom should be treated statistically on the same footing as other
degrees of freedom, which implies that the disorder average must be taken
over the sample partition function $ {\mathcal Z}[\rho(\vct r)]$ \cite{dotsenko1}.
Hence, the free energy of the annealed model reads
\begin{equation}
\label{eq:annealed_freen}
 \beta {\mathcal F}_{\mathrm{annealed}}
    = -  \ln\, \langle \!   \langle  {\mathcal Z}[\rho(\vct r)]  \rangle   \! \rangle.
\end{equation}
The annealed free energy may be evaluated using Eqs. (\ref{charge_correlation}), (\ref{eq:Z})
 and (\ref{eq:annealed_freen}) as
\begin{equation}
\beta{\mathcal F}_{\mathrm{annealed}}  = {\frac{1}{2}}\tr \ln
\big[G^{-1}({\vct r}, {\vct r}') + \beta  {\mathcal G}({\boldsymbol \varrho} - {\boldsymbol \varrho}'; z) \delta(z -z') \big].
                \label{eq:F_annealed}
\end{equation}
Note that unlike  the quenched case, Eq. (\ref{eq:F_quenched}),
the disorder contributions cannot be in general  separated from the pure
vdW contribution when the disorder is annealed.

In the case of two interacting planar dielectrics, we shall make use of Eqs. (\ref{eq:epsilon}) and (\ref{eq:bs_decomp})-(\ref{eq:g_b}) in order
to calculate the fluctuational trace-log of the modified inverse Green's function $G^{-1}({\vct r}, {\vct r}') + \beta   {\mathcal G}({\boldsymbol \varrho} - {\boldsymbol \varrho}'; z) \delta(z -z')$
in Eq. (\ref{eq:F_annealed}).  By employing
standard procedures \cite{rudi_jcp}, we find
\begin{equation}
\frac{\beta {\mathcal F}_{\mathrm{annealed}} }{S}  = { \frac{1}{2}
} \int \frac{{\mathrm{d}}^2Q}{(2\pi)^2} \ln
\big[1-\Delta_{1g}(Q)\Delta_{2g}(Q) e^{-2QD}\big],
\label{eq:F_fluct_charge_annealed}
\end{equation}
where we have (for $i=1, 2$)
\begin{eqnarray}
\Delta_{ig}(Q) &=&\\
&& \hspace{-1.4cm} = \frac{ \varepsilon_{m} Q \!-\!
\varepsilon_{i} \sqrt{Q^2 \!+\! 4\pi \ell_{\mathrm{B}} g_{ib} ~ c_{ib}
(Q)/\varepsilon_{i}  } \!-\! 4\pi \ell_{\mathrm{B}}  g_{is} ~\! c_{is} (Q) }
{ \varepsilon_{m} Q \!+\! \varepsilon_{i} \sqrt{Q^2 \!+\! 4\pi
\ell_{\mathrm{B}} g_{ib} ~ c_{ib} (Q)/\varepsilon_{i} } + 4\pi \ell_{\mathrm{B}}  g_{is}
~\! c_{is} (Q)}.\nonumber
\label{delta_i_g_Quenched} 
\end{eqnarray}
The total force, $f_{\mathrm{annealed}} = -\partial {\mathcal
F}_{\mathrm{annealed}}/\partial D$, in the annealed model thus
follows as
\begin{equation}
\frac{\beta {f}_{\mathrm{annealed}} }{S}  = - \frac{1}{2 \pi} \!\int
Q^2 {\mathrm{d}} Q  \frac{\Delta_{1g}(Q) \Delta_{2g}(Q)\, e^{-2QD}}{1 -
\Delta_{1g}(Q) \Delta_{2g}(Q)\,e^{-2QD}}.
\label{eq:force_annealed}
\end{equation}
The annealed interaction free energy, or the corresponding force, has a form reminiscent
of the zero-frequency vdW interaction between two semi-infinite slabs \cite{ninham} with renormalized
dielectric missmatch, $\Delta_{ig}(Q)$, depending on the in-plane wave-vector $Q$. Indeed, one can 
argue for the following exact correspondence based on the vdW interactions between media with volume and 
surface embedded mobile charges \cite{ninham,rudiold}: the term $\sqrt{Q^2 \!+\! 4\pi \ell_{\mathrm{B}} g_{ib} ~ c_{ib}
(Q)/\varepsilon_{i}  }$ corresponds to bulk Debye-like  ``screening", stemming from the annealed bulk disorder response
to local electrostatic fields. It can be obtained alternatively by analyzing the fluctuational modes of the 
Debye-H\"uckel fields as opposed to the Laplace fields \cite{ninham}. The term $4\pi \ell_{\mathrm{B}}  g_{is}~\! c_{is} (Q)$, 
on the other hand, stemms from the response of the mobile charges confined to the dielectric surfaces to electrostatic fields, 
corresponding to surface disorder-generated  ``screening" \cite{rudiold}. It can be derived alternatively by considering the 
response of the surface polarization to local electrostatic fileds \cite{rudiold}. The combination of the two terms in 
Eq. (\ref{delta_i_g_Quenched}) can not be written as a sum of two terms depending linearly on $c_{ib}$ and $c_{is}$. 
This signifies that the bulk and surface disorder effects for the annealed case are obviously not additive and can not be analyzed separately.

\begin{figure*}[t]\begin{center}
	\begin{minipage}[b]{0.35\textwidth}\begin{center}
		\includegraphics[width=\textwidth]{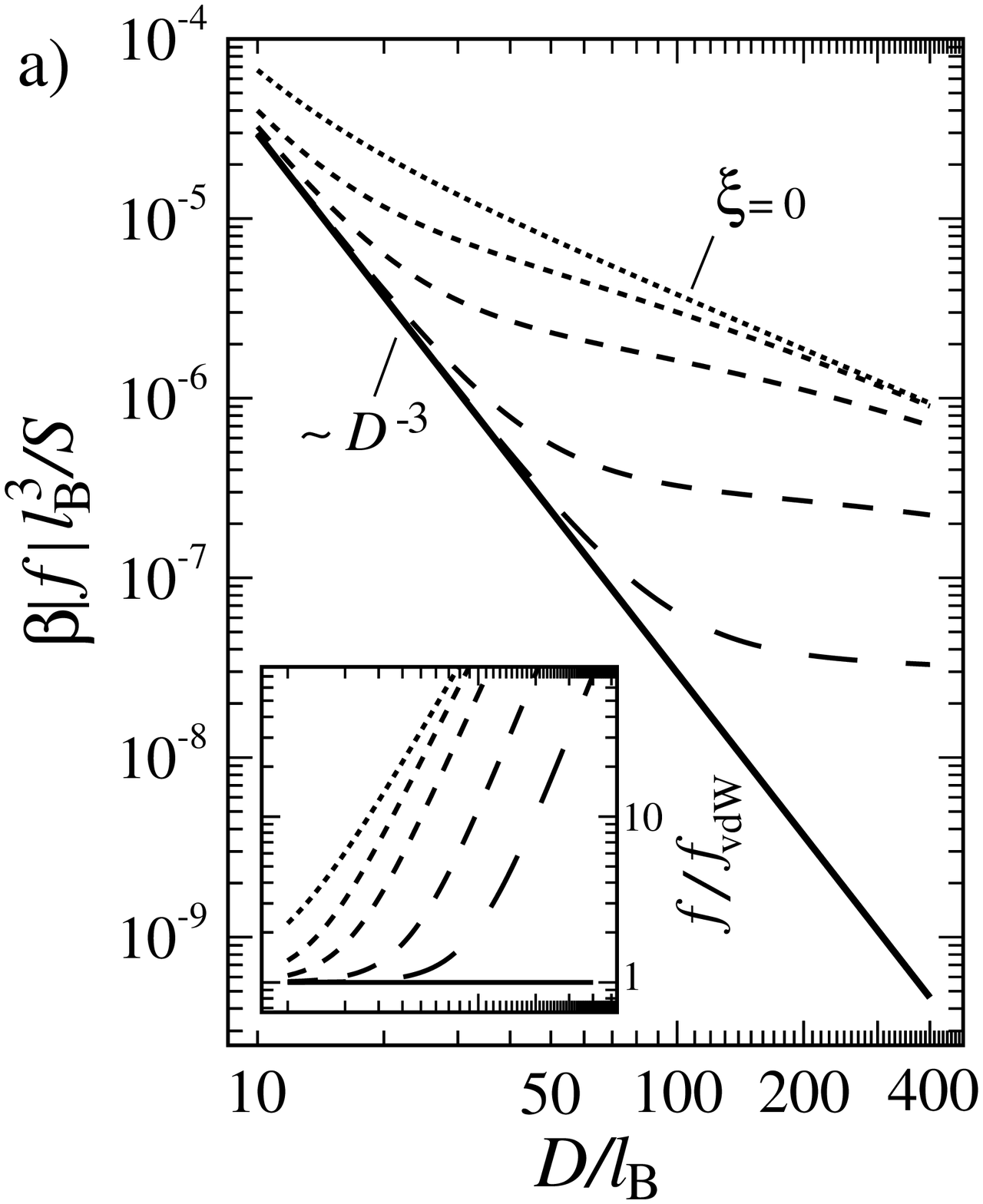} 
	\end{center}\end{minipage} \hskip-1cm
	\begin{minipage}[b]{0.35\textwidth}\begin{center}
		\includegraphics[width=\textwidth]{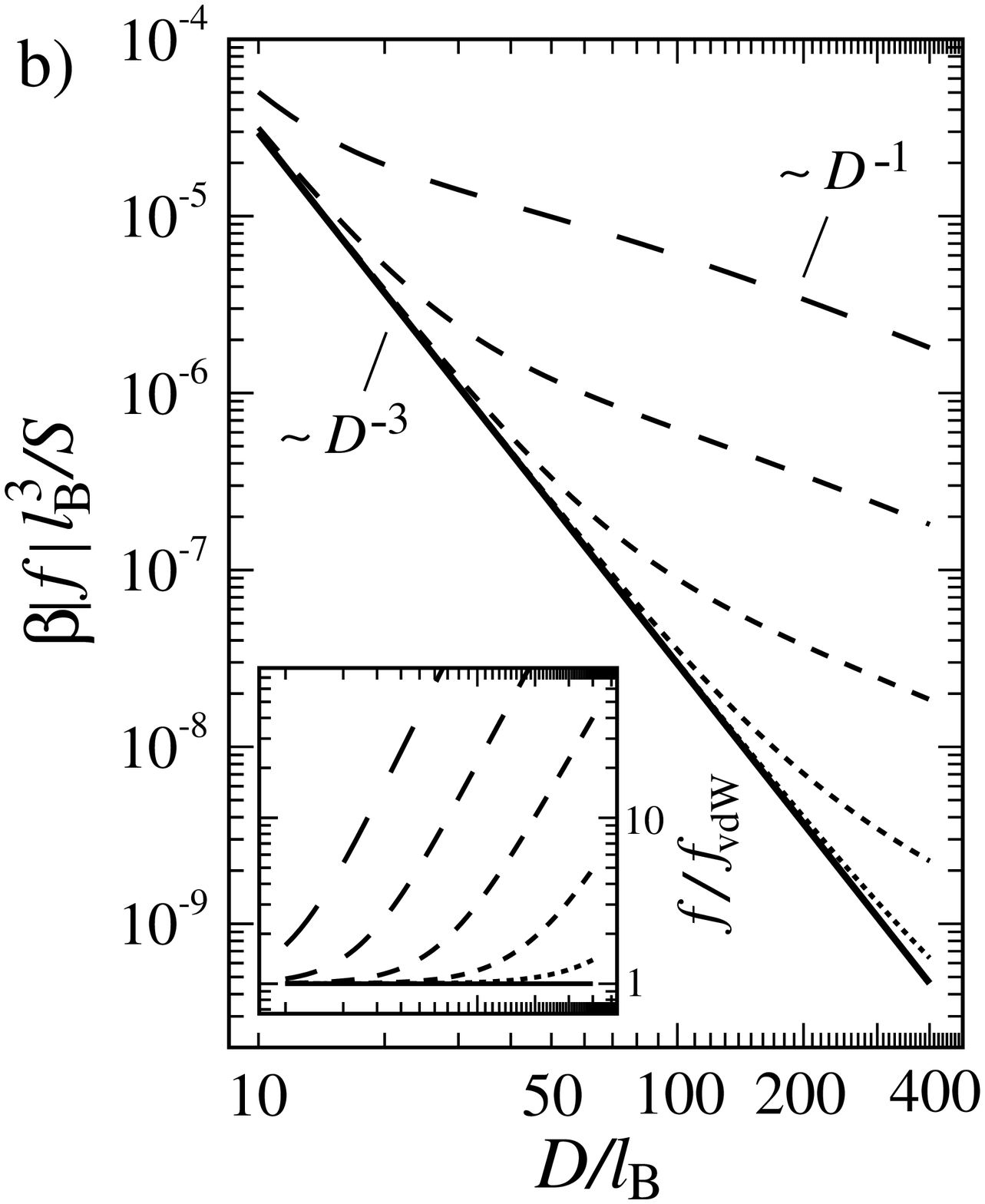} 
	\end{center}\end{minipage} \hskip-1cm
	\begin{minipage}[b]{0.35\textwidth}\begin{center}
		\includegraphics[width=\textwidth]{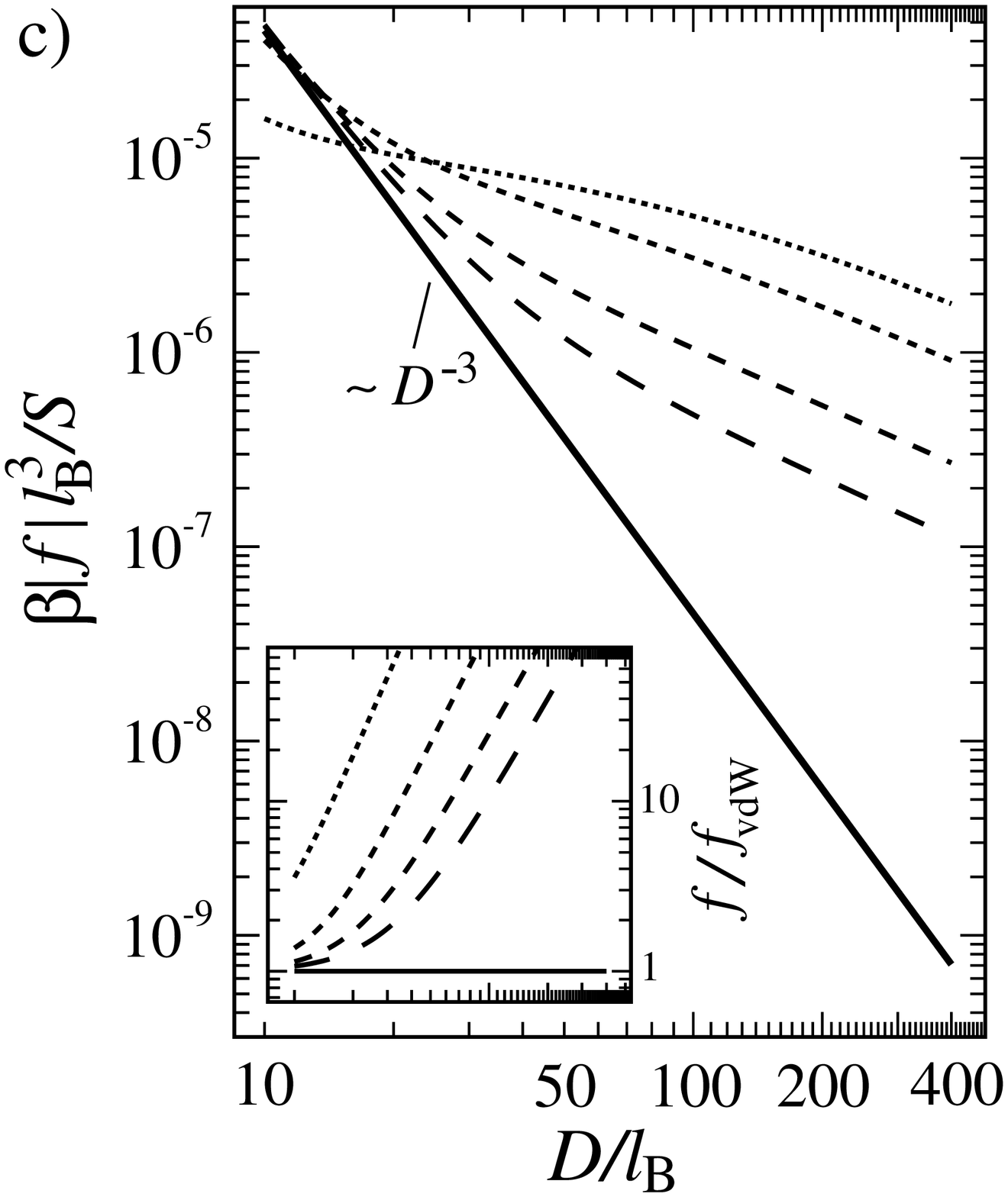}
	\end{center}\end{minipage} \hskip-1cm	
\caption{a) Magnitude of the rescaled total force, $\beta |f|
l_{\mathrm{B}}^3 /S$ (Eq. (\ref{eq:force_quenched})), between
two identical net-neutral dielectric slabs in {\em vacuum}
($\varepsilon_m =1$) bearing {\em quenched} monopolar charge
disorder as a function of the rescaled distance, $D/l_{\mathrm{B}}
$. The results are plotted here for fixed $\varepsilon_p=10$,
$g_s=0$ (no surface disorder) and bulk disorder variance $g_b = 5
\times 10^{-8}\, {\mathrm{nm}}^{-3}$ and varying disorder
correlation length  $\xi / l_{\mathrm{B}} =
0 , 200, 10^3, 10^4, 10^5$ (from top to bottom). Inset shows
the ratio of the total force to the pure zero-frequency vdW force
(\ref{eq:vdW}) between the slabs in the absence of charge disorder
for the same range of $D$. b) Same as (a) but here we fix
$\varepsilon_p=10$,  $ g_s = 0$, $\xi / l_{\mathrm{B}} =200 $ and vary the
bulk disorder variance in the range $g_b = 10^{-7},
10^{-8}, 10^{-9}, 10^{-10}, 10^{-11}\, {\mathrm{nm}}^{-3}$ (from top
to bottom). Solid curve is the pure vdW force (\ref{eq:vdW}). c)
Same as (b) but here we fix $\xi / l_{\mathrm{B}} = 200$,  $g_b = 5
\times 10^{-8}\, {\mathrm{nm}}^{-3}$,  $g_s = g_b^{2/3} $ and
vary the dielectric constant of slabs as $\varepsilon_p = 2$ (top
dotted curve), 10,   40, 100 (bottom dot-dashed curve). All
graphs are plotted in log-log scale. }
\label{fig:fig2}
\end{center}
\end{figure*}

\section{Symmetric case of two identical slabs}
\label{sec:symmetric}

We now analyze the preceding analytical results in the situation where the two dielectric slabs
are identical, i.e., we have $\varepsilon_i= \varepsilon_p$  ($\Delta_i = \Delta$),
$g_{is} = g_s$, $g_{ib} = g_b$, and $\xi_{is}= \xi_{ib} = \xi$ for both slabs $i=1, 2$.

\subsection{Quenched disorder-induced interactions}
\label{sec:symm_quenched}

In the case of quenched disorder, one can expand the Lorentzian correlation function $c_{i\alpha}(Q)$, Eq. (\ref{eq:Fourier_correlation_function}),
in powers of  the dimensionless ratio of the correlation length to the intersurface separation, $\xi/D$, and thus express the total
force (\ref{eq:force_quenched}) in the form of a series expansion as
\begin{equation}
f_{\mathrm{quenched}} =  f^{(0)} + \sum_{n=1}^{\infty} f^{(n)},
\label{eq:f0_fn}
\end{equation}
where we have
\begin{equation}
 \frac{ \beta {f}^{(0)}}{S} = - \frac{g_b \ell_{\mathrm{B}}
\Delta }{2 \varepsilon_p D} - \frac{2 \varepsilon_m g_s \ell_{\mathrm{B}}
\ln (|1-\Delta^2|) }{(\varepsilon_m+\varepsilon_p)^2 \Delta \,D^2} - \frac{{\mathrm{Li}}_3(\Delta^2)}{8\pi D^3},
\label{eq:f_0_quenched}
\end{equation}
corresponding to the free energy of the system in the presence of
a completely uncorrelated disorder ($\xi=0$), and
\begin{eqnarray}
\frac{ \beta {f}^{(n)}}{S} = \!& -& \!\frac{4 g_b \ell_{\mathrm{B}}
\varepsilon_m}{(\varepsilon_m+\varepsilon_p)^2 \Delta D}
\!\sum_{n=1}^{\infty} (-1)^n \frac{\xi^{2n}}{D^{2n}} C_{2n+1} \mathrm{Li}_{2 n} (\Delta^2) \nonumber\\
&& \hspace{-1.9cm} - \frac{8 g_s \ell_{\mathrm{B}}
\varepsilon_m}{(\varepsilon_m + \varepsilon_p)^2 \Delta D^2}
\!\sum_{n=1}^{\infty} (-1)^n \frac{\xi^{2n}}{D^{2n}} C_{2n+2} \mathrm{Li}_{2 n+1} (\Delta^2),
\label{eq:f_n_quenched}
\end{eqnarray}
with $C_n=\Gamma (n)/2^n$.
The above series expansion is most suitable for
the situation where $\xi/D$ is small.

Let us first consider the leading-order uncorrelated disorder free energy
${f}^{(0)}$ in Eq. (\ref{eq:f_0_quenched}). This contribution exhibits a few remarkable
features \cite{PRL}. First, it shows  a sequence of scaling behaviors with
the separation that stem from  different origins:
a leading $1/D$ term due to the quenched bulk
disorder, a subleading $1/D^2$ term  from the surface charge disorder, and the pure
vdW term that goes as $1/D^3$, i.e.
\begin{equation}
\frac{\beta f_{{\text {vdW}}}}{S} =
-\frac{{\mathrm{Li}}_3(\Delta^2)}{8\pi D^3}.
\label{eq:vdW}
\end{equation}
 While the vdW term is always attractive in a symmetric system, the
disorder contributions (first and second terms in Eq.
(\ref{eq:f_0_quenched})) are attractive when the dielectric
mismatch $\Delta>0$ (i.e., when the dielectric constant of the
intervening medium is smaller than that of the slabs) and
repulsive otherwise. The former case ($\Delta>0$) has been
investigated in detail previously in the context of  two identical
dielectric slabs in vacuum ($\varepsilon_m = 1$) \cite{PRL}. 

Note that since the forces induced by an uncorrelated disorder
exhibit a much weaker decay with the separation, they may
completely mask the standard vdW force (which is always present
regardless of any charge disorder) at sufficiently large
separations \cite{PRL}. One might expect that globally
electroneutral slabs would exhibit a dipolar-like interaction
force to the leading order rather than the monopolar forms $1/D$
(or $1/D^2$) obtained for the bulk (or surface) charge
distribution. As noted before, the physics involved is indeed subtle as the
disorder terms result from  the self-interaction of the charges
with their images (which follows from $G ({\mathbf r}, {\mathbf r})$, Eq.
(\ref{eq:F_quenched}), in the limit of zero correlation length) and not from
dipolar interactions (which come from an expansion of $G({\mathbf r}, {\mathbf r}')$
when $|{\mathbf r} - {\mathbf r}'|$ is large). Statistically speaking each charge on
average (as any other charge has an equal probability of being of
the same or opposite sign) only sees its image, thus explaining
the leading monopolar form in the net force.


The next leading correction due to disorder correlations for
small  $\xi / D$ follows from  Eq. (\ref{eq:f_n_quenched}) as
\begin{equation}
\frac{ \beta {f}^{(1)}}{S} \simeq \frac{g_b \ell_{\mathrm{B}} \varepsilon_m  \xi^2 \Delta\,
\mathrm{Li}_2 (\Delta^2)}{(\varepsilon_p + \varepsilon_m)^2
D^3} +\frac{3 \varepsilon_m g_s \ell_{\mathrm{B}} \xi^2\, \mathrm{Li}_3
(\Delta^2)}{(\varepsilon_m + \varepsilon_p)^2 \Delta \,D^4},
\label{eq:force_quenched_total}
\end{equation}
which has an opposite sign to the zeroth-order term, and thus
tends to weaken the zeroth-order effect. Note however that the higher-order corrections in Eq. (\ref{eq:f_n_quenched}) alternatively change sign
and their net effect leads to a total force which can be evaluated
numerically, e.g., directly
from Eq. (\ref{eq:force_quenched}). The results are  shown
in Fig. \ref{fig:fig2} for the case of two identical dielectric slabs
in vacuum. As seen in Fig. \ref{fig:fig2}a,
the attractive disorder-inducd forces indeed
diminish (albeit rather slowly as seen in the inset) as
the disorder correlation length is increased (by several
orders of magnitude from $\xi=0$ up to $\xi\simeq 5$mm
in actual units in the figure). For a large correlation length
or at small separation (large $\xi / D$), the total force thus tends
to the non-disoredred  vdW force (\ref{eq:vdW}) that scales as $1/D^3$
(thick solid line), while for a small correlation length or
at large separation (small $\xi / D$), the force increases and shows a
crossover to the maximal uncorrelated disorder value (\ref{eq:f_0_quenched}) that scales as $1/D$
(top dotted line). The latter is obviously from the bulk disorder which, if
present, gives rise to the most dominant effects.

The crossover behavior depends also significantly on the disorder
variance and the dielectric mismatch (Figs. \ref{fig:fig2}b and c).
Similar trends as described above can be observed by varying the bulk disorder 
variance, $g_b$, as shown in Fig. \ref{fig:fig2}b (here $g_b$ varies within  the typical range $10^{-11}-10^{-7}\,{\mathrm{nm}}^{-3}$ corresponding to impurity charge densities of
$10^{10}-10^{14}\,e_0/{\mathrm{cm}}^{3}$ \cite{Kao_Pitaevskii}).
The results in Fig. \ref{fig:fig2}c however reveal
a nonmonotonic dependence on the dielectric mismatch:
The disorder contribution tends to decrease both for small and large
slab dielectric constant. This can be seen directly from
Eqs.  (\ref{eq:f_0_quenched}) and (\ref{eq:f_n_quenched})
as the disorder-induced force vanishes
in the limits $\varepsilon_p\rightarrow \infty$ or  $\Delta\rightarrow 1$ (perfect conductor)
and $\Delta\rightarrow 0$ (homogeneous medium, in which case the vdW
force vanishes as well).

It is interesting to note that
the first-order corrections due to {\em bulk correlations}
scales as $1/D^3$, which is similar to the vdW contribution. Hence, these
corrections tend to renormalize the ideal zero-frequency
Hamaker coefficient, $A \equiv 3(k_{\mathrm{B}}T){\mathrm{Li}}_3(\Delta^2)/4$,
associated with the vdW term (defined via
$ f_{\mathrm{vdW}}/S =  -  A/(6\pi D^3)$) to an effective value
given by
\begin{equation}
   A_{\mathrm{eff}}  = \frac{3k_{\mathrm{B}}T}{4}{\mathrm{Li}}_3(\Delta^2) +  \frac{3g_b e_0^2 \,\varepsilon_m \Delta \xi^2
\mathrm{Li}_2 (\Delta^2)}{2\varepsilon_0(\varepsilon_p + \varepsilon_m)^2 }.
\end{equation}
This value is larger (smaller) than the ideal value $A$ when $\Delta>0$
($\Delta<0$) and can even change sign.

\begin{figure}[t!]\begin{center}
\vskip-1.9cm
\begin{minipage}[b]{0.36\textwidth}\begin{center}
\includegraphics[width=\textwidth]{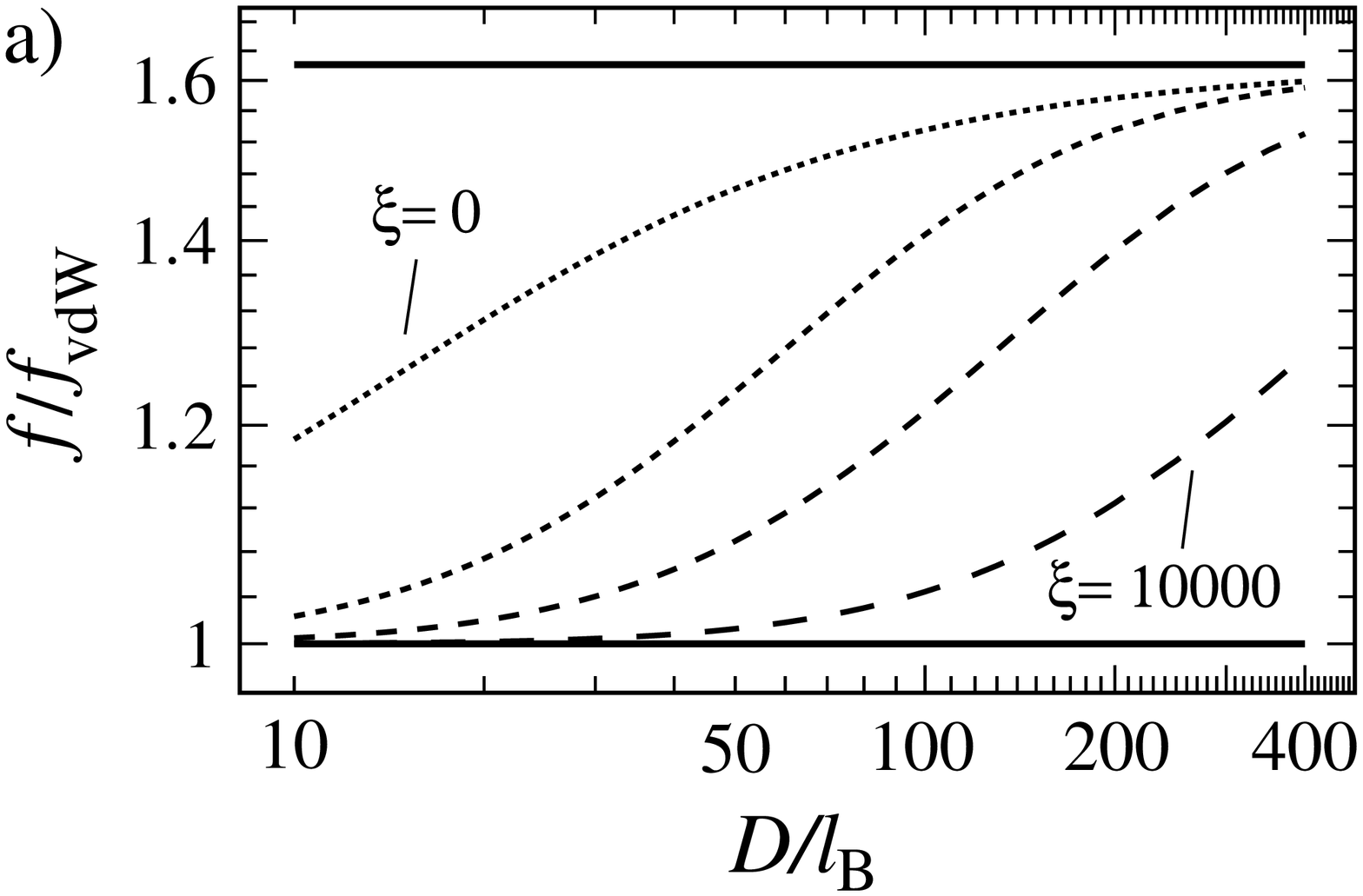} 
\end{center}\end{minipage} 
\vskip-1.9cm
\begin{minipage}[b]{0.36\textwidth}\begin{center}
\includegraphics[width=\textwidth]{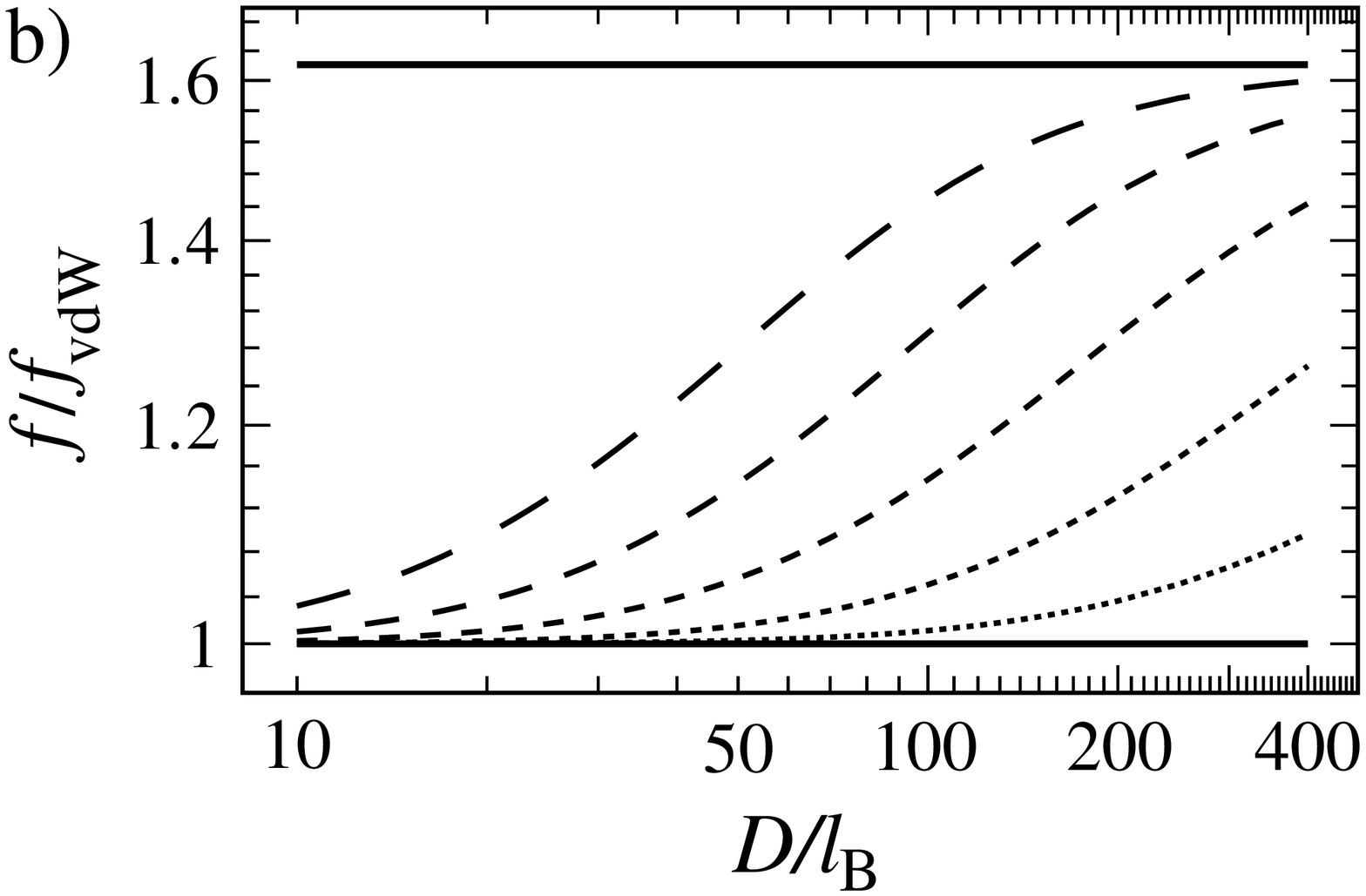} 
\end{center}\end{minipage}
\vskip-1.4cm
\begin{minipage}[b]{0.36\textwidth}\begin{center}
\includegraphics[width=\textwidth]{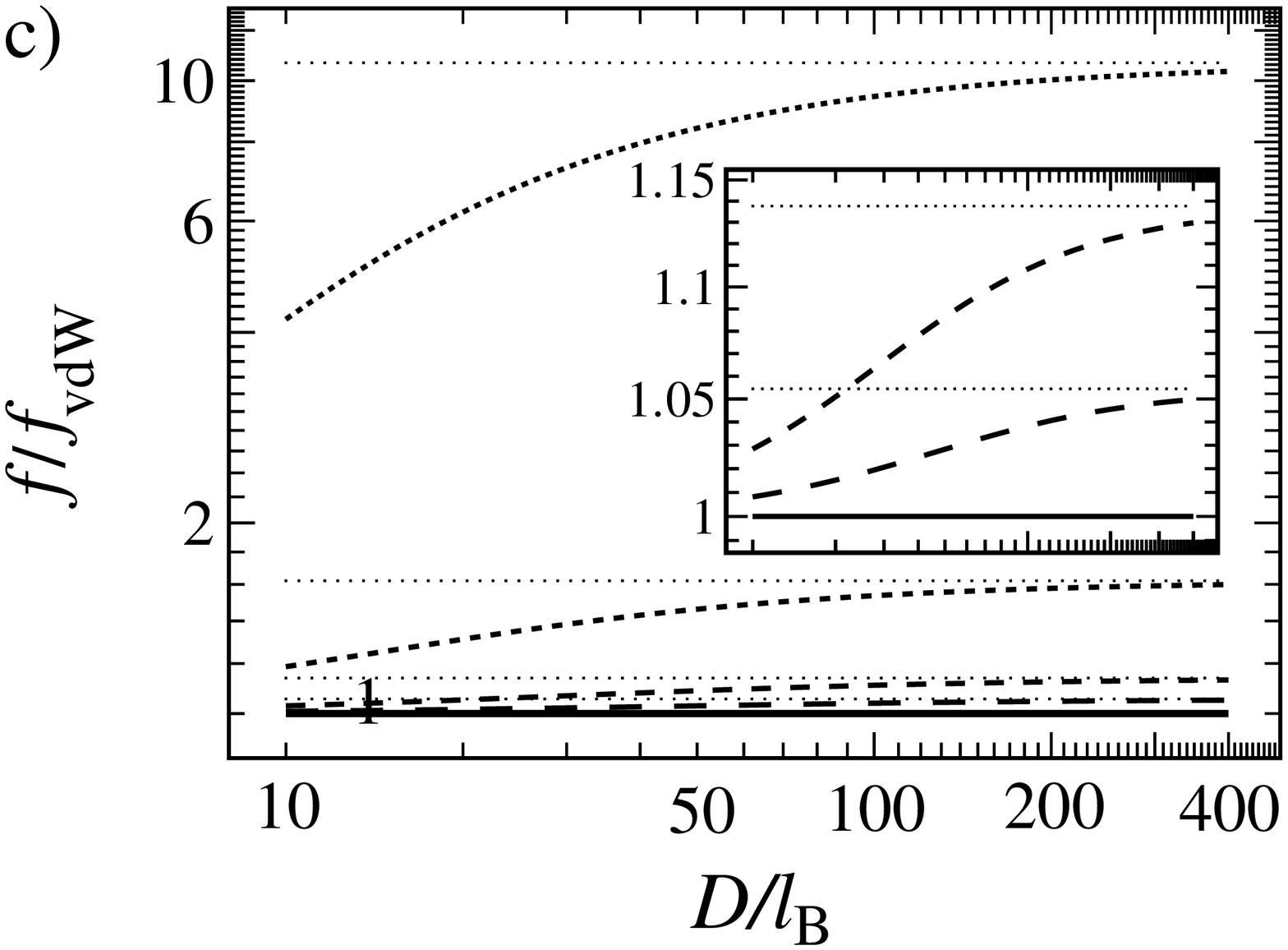} 
\end{center}\end{minipage}
\caption{ a)
Ratio of the total force (\ref{eq:force_annealed}) to the
zero-frequency vdW force (\ref{eq:vdW}) between two identical
net-neutral dielectric slabs in {\em vacuum} ($\varepsilon_m =1$)
bearing {\em annealed} monopolar charge disorder as a function of
the rescaled distance, $D/ l_{\mathrm{B}}$. The results are
plotted here for fixed $\varepsilon_p =10$, $g_s =0$ (no surface
disorder), $ g_b
= 5 \times 10^{-8}\, {\mathrm{nm}}^{-3}$ and varying  disorder
correlation length  $\xi / l_{\mathrm{B}}= 0, 200,
10^3, 10^4 $ (from top to bottom).
Annealed curves are bounded by the perfect conductor result, Eq.
(\ref{eq:metal}) (top solid line), for large disorder and the vdW
result, Eq. (\ref{eq:vdW}) (bottom solid line), for no disorder.
b) Same as (a) but here we have fixed 
$\varepsilon_p =10$, $g_s =0$, $\xi / l_{\mathrm{B}} =200$ 
and bulk disorder variance
varied in the range $g_b = 10^{-7} , 10^{-8}, 10^{-9},
10^{-10}, 10^{-11}\, {\mathrm{nm}}^{-3}$ (from top to bottom).
 c) Same as (a) but here we fix
$g_s =0$, $g_b = 5 \times 10^{-8}\, {\mathrm{nm}}^{-3}, \xi =0
$, and vary the dielectric constant of the slabs as $\varepsilon_p
= 2, 10, 40, 100$ (from top). Inset shows a closer view of
the curves for $\varepsilon_p =40, 100$ (from top). Top 
dotted lines correspond to Eq. (\ref{eq:metal}). All graphs are
plotted in log-log scale.} 
\label{fig:fig3}
\end{center}
\end{figure}
%

\subsection{Annealed disorder-induced interactions}

In the case of annealed disorder, the total force between the dielectric slabs is given by Eq. (\ref{eq:force_annealed}).
It shows that  the force in this case cannot be expressed simply as a sum of different
additive contributions as in the quenched case. Rather, it follows from
the combined effect of  the annealed disorder fluctuations and the underlying vdW effect, which again 
reflects the fact that the disorder charge distribution in  this case
can adapt itself in order to minimize the total free energy of the system.
\begin{figure}[t!]\begin{center}
\vskip-1.4cm
\begin{minipage}[b]{0.36\textwidth}\begin{center}
\includegraphics[width=\textwidth]{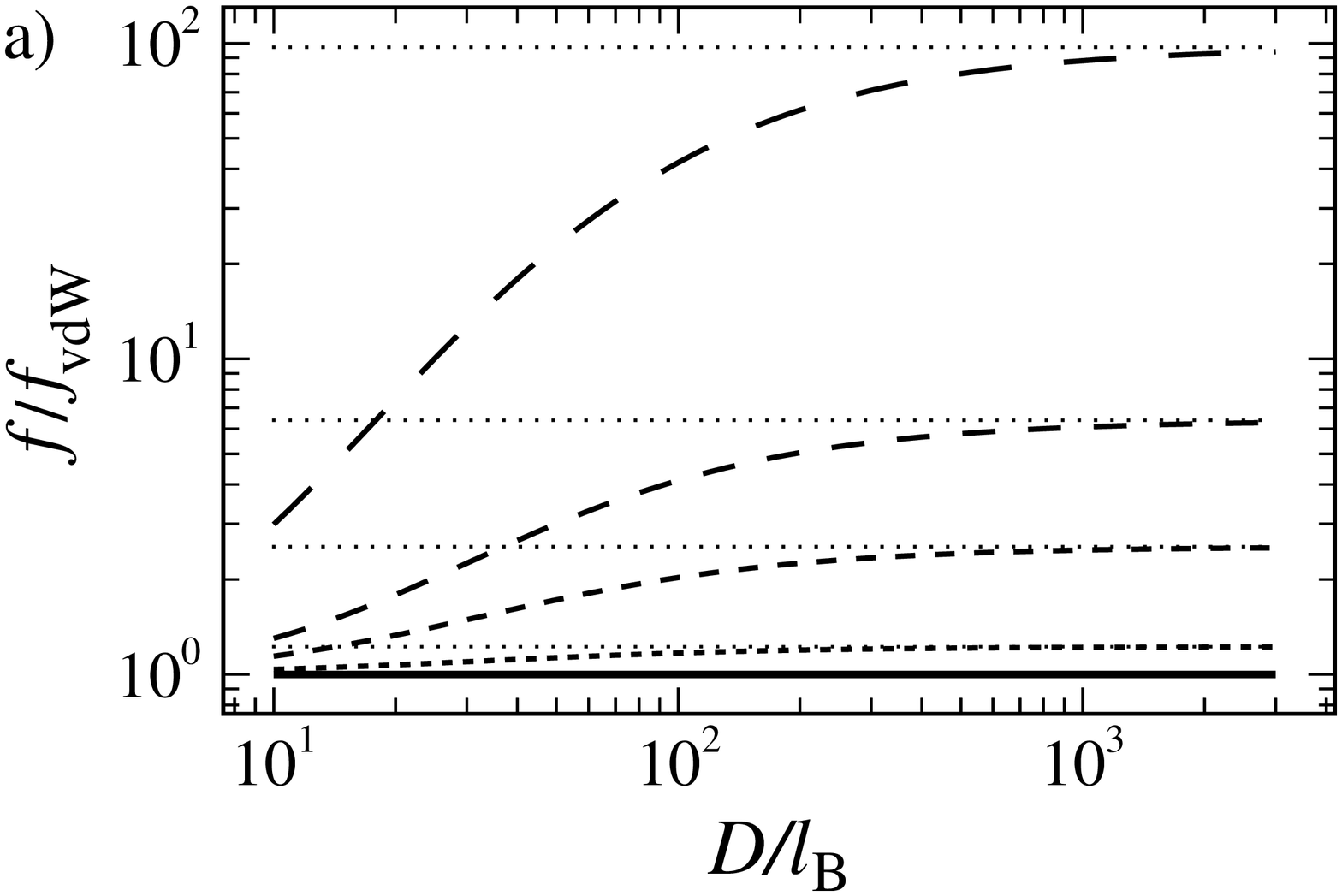} 
\end{center}\end{minipage} 
\vskip-1.7cm
\begin{minipage}[b]{0.36\textwidth}\begin{center}
\includegraphics[width=\textwidth]{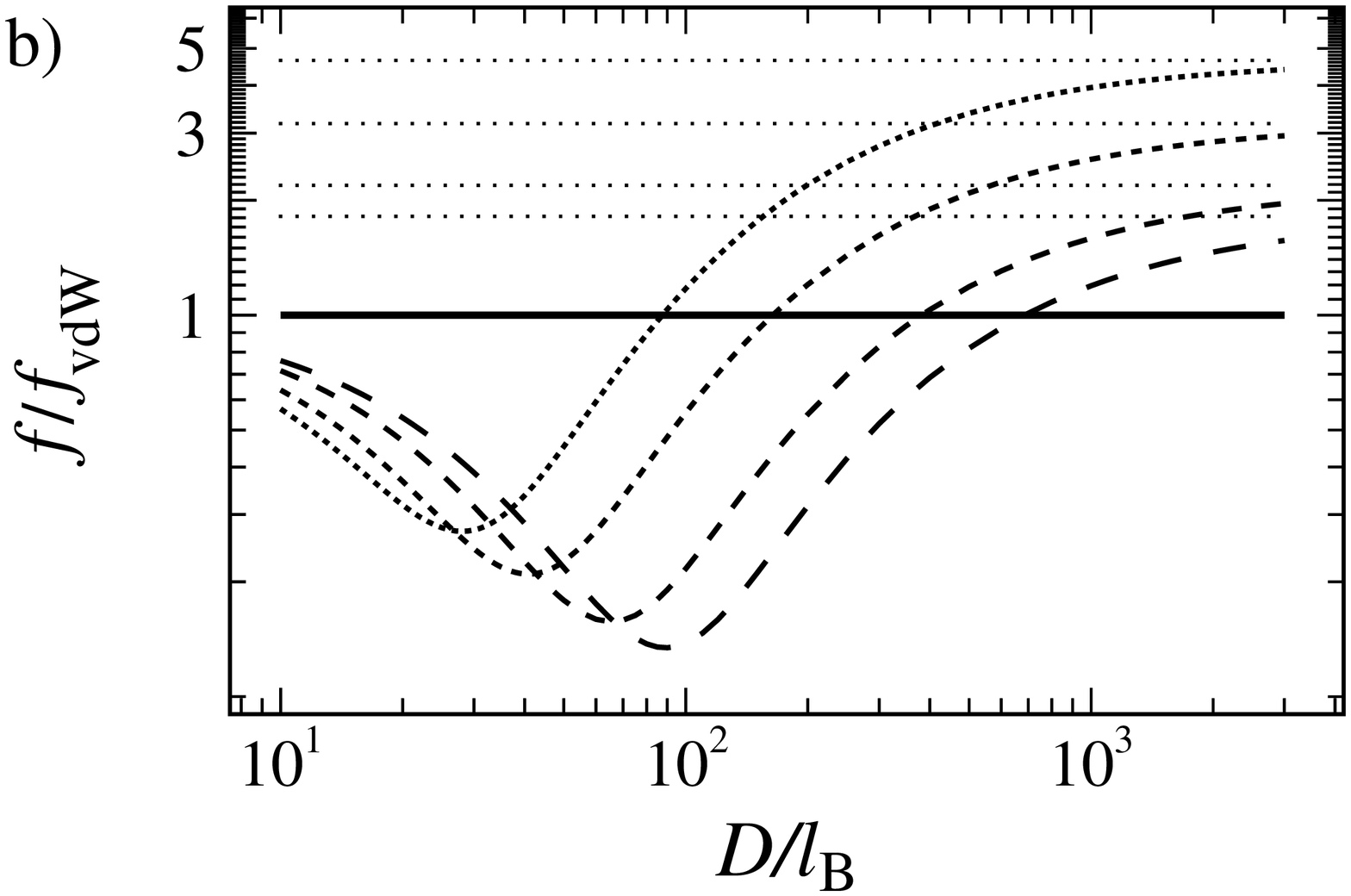} 
\end{center}\end{minipage}
 \caption{a) Same
as Fig. \ref{fig:fig3}a but here we fix $\varepsilon_p =50$, $g_s
=0$, $g_b= 5 \times 10^{-8}\, {\mathrm{nm}}^{-3}$, $\xi =0$, and vary
the dielectric constant of the intervening medium as
$\varepsilon_m =2, 10, 20, 40$ (from bottom to top). b) Same as
(a) but here we fix the dielectric constant of the slab as
$\varepsilon_p =10$ and vary that of the intervening medium as
$\varepsilon_m =30, 40 , 60, 80$ (from top to bottom). Top 
dotted lines correspond to Eq. (\ref{eq:metal}). 
} \label{fig:fig4}
\end{center}
\end{figure}

The resulting total force is shown in Fig. \ref{fig:fig3} for the
symmetric case of two identical dielectric slabs in vacuum. For
the sake of presentation, we have plotted the ratio of the
annealed force  (\ref{eq:force_annealed}) to the vdW force
(\ref{eq:vdW}), which clearly shows that the net force is
bounded by two {\em limiting laws}, namely,  the ideal limiting force
\begin{equation}
    \frac{\beta f_{\mathrm{ideal}} }{S} = -\frac{\zeta(3)}{8\pi D^3},
        \label{eq:metal}
\end{equation}
which is similar to the one obtained between perfect conductors
\cite{bordag}, and the vdW force (\ref{eq:vdW}) in the absence of
any disorder charges. These constitute the upper and the lower
bounds for the annealed force in the general symmetric case with
$\Delta
> 0$, i.e.
%
\begin{equation}
|f_{\mathrm{vdW}}|   < |f_{\mathrm{annealed}}| <  |f_{\mathrm{ideal}}|.
\end{equation}
Note that the above limiting values can be established
systematically from the general expression
(\ref{eq:force_annealed}), and are strictly attractive (negative
sign). The lower bound vdW force is non-universal (i.e., material
dependent) and is obtained asymptotically in the limit of small
disorder variance ($g_b$, $g_s\rightarrow 0$) or small separationss.
The upper bound is {\em universal} and follows in the limit of
strong disorder variance ($g_b$ or $g_s\rightarrow \infty$) or
large separations. The crossover from  one limit to the other can
be achieved by increasing the distance or the disorder variance as
seen in Fig. \ref{fig:fig3}.

In the annealed case, the disorder correlations play a similar
role as in the quenched case and tend to weaken the
disorder-induced forces. As seen in Fig. \ref{fig:fig3}a, the
magnitude of the force drops from its  value for an uncorrelated
disorder ($\xi=0$), and also the crossover between the two
limiting behaviors as discussed above is  `delayed' when the
disorder has a finite correlation length, $\xi$.

\begin{figure}[t]
\vspace{-2cm}
\includegraphics[angle=0,width=7.1cm]{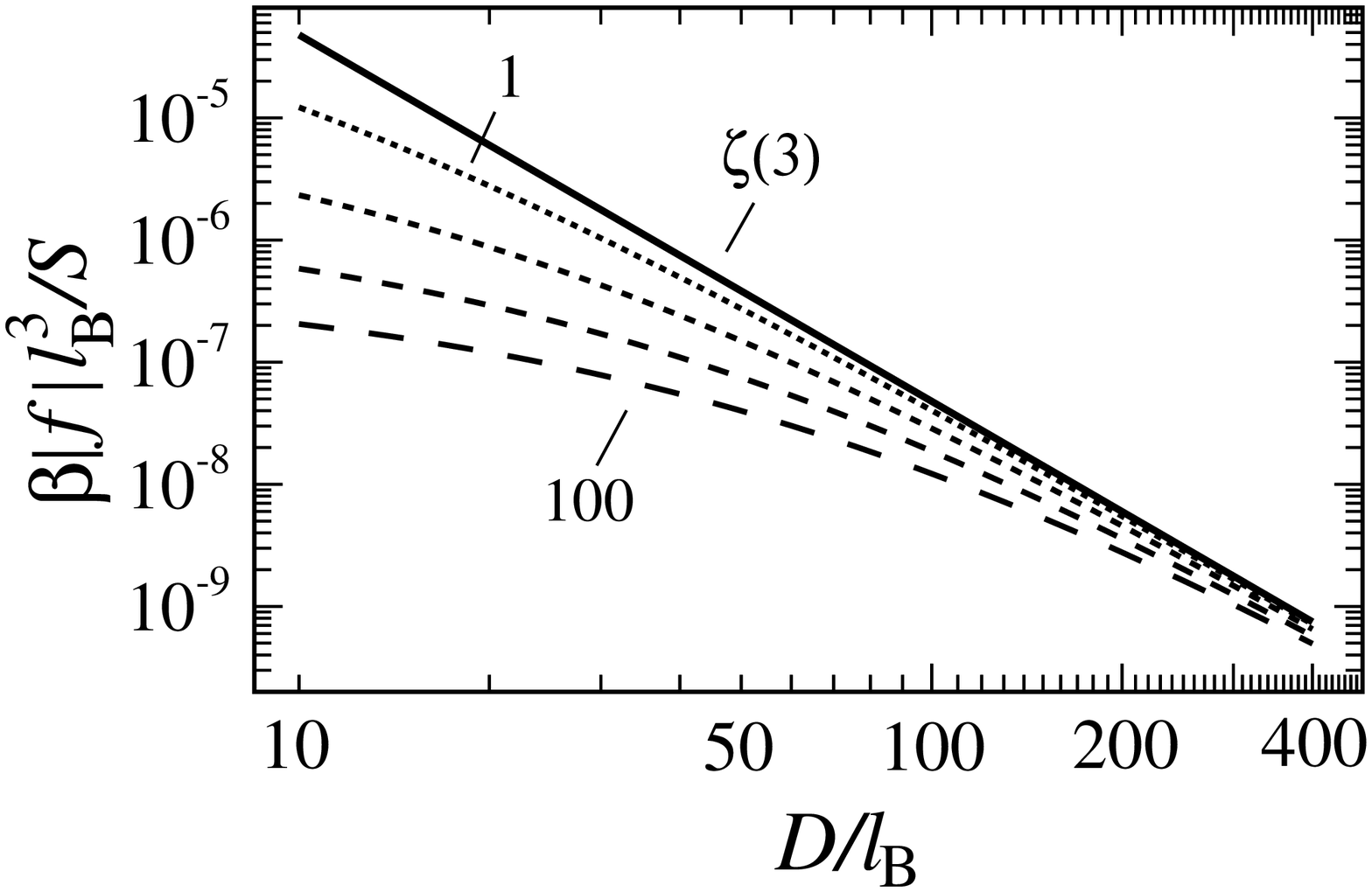}
\caption{The
rescaled magnitude of the (attractive) force, Eq.
(\ref{eq:force_annealed}), between two identical net-neutral
dielectric slabs bearing annealed monopolar charge disorder in a
{\em dielectrically homogeneous} system ($\varepsilon_1
=\varepsilon_2 = \varepsilon_m$) as a function of the rescaled
distance, $D / l_{\mathrm{B}}$. Here we plot the resulting force
for $\varepsilon_1 =\varepsilon_2 =
\varepsilon_m=1,10,40,100$ (from top to bottom) and for
uncorrelated bulk disorder in both slabs with $\xi =0$, $g_b = 5
\times 10^{-8} \,{\mathrm{nm}}^{-3}$ and $g_s =0$. Top solid line shows
the universal limiting expression (\ref{eq:metal}). } \label{fig:fig5}
\end{figure}

It is also remarkable to note that Eq. (\ref{eq:metal}) is obtained in general as the large-distance
($D\rightarrow \infty$) behavior for the total force between any two {\em arbitrary} dielectric slabs
regardless of their dielectric constant and disorder variance. It thus  demonstrates
the intuitive fact that dielectric slabs with annealed charges tend to behave asymptotically in a way similar to perfect conductors.
This is one of the distinctive features of the annealed disorder as compared with the quenched disorder,
whose effects depend significantly on the material properties. The deviations due to material properties and
the disorder variance in the annealed case contribute a repulsive
subleading force in the symmetric case, which can be determined from a series
expansion at large separations.
If the surface disorder is negligible as compared with the bulk disorder ($g_s \ll g_b
\ell_{\mathrm{B}}$),  we obtain (up to the first few leading orders)
\begin{eqnarray}
\frac{\beta f_{\mathrm{annealed}} }{S} \simeq \!\!&-&\!\!
\cfrac{\zeta(3)}{8\pi D^3} \!+\! \cfrac{ 3 \varepsilon_m
\zeta(3)}{\sqrt{ 64\pi^3 g_b \ell_{\mathrm{B}} \varepsilon_p}D^4} - \frac{3
\varepsilon_m^2 \zeta (3)}{8 \pi^2 g_b \ell_{\mathrm{B}} \varepsilon_p D^5}
\nonumber\\
&+& \!\! \frac{15 \xi^2 \varepsilon_m \zeta (5)}{2 \sqrt{64 \pi^3
g_b \ell_{\mathrm{B}} \varepsilon_p}D^6} \! - \! \frac{45 \xi^2
\varepsilon_m^2 \zeta (5)}{16 \pi^2 g_b \ell_{\mathrm{B}} \varepsilon_p D^7}.
\label{eq:total_f_annealed_1}
\end{eqnarray}
While in the opposite situation where the bulk disorder is negligible ($g_s \gg g_b \ell_{\mathrm{B}}$),  we obtain
\begin{equation}
\frac{\beta f_{\mathrm{annealed}} }{S} \simeq
-\cfrac{\zeta(3)}{8\pi D^3} + \cfrac{3 \varepsilon_m
\zeta(3)}{16\pi^2 g_s\ell_{\mathrm{B}} D^4} + \frac{15
\varepsilon_m \xi^2 \zeta (5)}{16 \pi^2 \ell_{\mathrm{B}} g_s
D^6}.
\label{eq:total_f_annealed_2}
\end{equation}

We should emphasize that, although the annealed force decays at
large separations in a similar fashion as the pure vdW force, its
magnitude can nevertheless exceed the vdW force by a few orders of
magnitude if the dielectric constant of the intervening material
$\varepsilon_m$ is increased toward that of the slabs (see Fig.
\ref{fig:fig4}a). 
The asymptotic behavior of the annealed force is similarly
described by the pure vdW result (\ref{eq:vdW}) and the ideal
expression (\ref{eq:metal}) even if the dielectric constant of the
(identical) slabs is smaller than that of the intervening medium,
$\Delta <0$. However, in this case, these two limiting results do
not constitute the upper and lower bound limits for the total
force as the force ratio $f/ f_{\mathrm{\text{vdW}}}$
exhibits a nonmonotonic behavior with  a local minimum at some
intermediate separation between the slabs (Fig.
\ref{fig:fig4}b).

The special case of a dielectrically homogeneous
system, $\varepsilon_1=\varepsilon_2=\varepsilon_m$, constitutes
another example where annealed and quenched disorder effects
differ on a qualitative level and may thus be easily
distinguished. In this case, the total force
(\ref{eq:force_quenched}) due to quenched disorder vanishes
trivially at all separations, while the total force due to
annealed disorder remains finite. Since the vdW contribution
vanishes in a dielectrically  homogeneous system as well, the total
force in this case comes purely 
from the electrostatic interactions of annealed charges in the two
slabs (Fig. \ref{fig:fig5}). 

\section{Asymmetric case of two dissimilar slabs}
\label{sec:asymmetric}

So far we have considered only the case of a symmetric
system composed of two identical semi-infinite slabs. 
In practice, however, one may often deal with a situation
where the dielectric constant or disorder variance of the two slabs  are 
different. In this case, the resulting
fluctuation-induced interactions may exhibit qualitatively different features as compared with the fully symmetric case that we shall explore further in this 
Section.  

\subsection{Interaction between a disordered and a disorder-free slab}

Let us first consider briefly the situation where the two slabs are
dielectrically identical, $\varepsilon_1=\varepsilon_2$ ($\Delta_1=\Delta_2=\Delta$), but bear different
degrees of quenched or annealed monopolar charge disorder. For the sake of simplicity, let us
assume that one slab  is {\em disorder free} ($g_{1b}, g_{1s}=0$ and $\xi_1=0$),  whereas the other slab
contains disorder charges of bulk and surface variance $g_{2b}$ and $g_{2s}$, and correlation
lengths $\xi_{2b}=\xi_{2s}=\xi_2$ (recall that in any case
the net monopolar charge in each slab is taken to be zero).

This case is particularly interesting because
it shows the interconnection between the disorder and the dielectric inhomogeneity in the system.
In the quenched case, we find that the disorder contribution to the total force is nonzero (even though one of the slabs does not
contain any disorder charges), and is simply given by {\em half} the value obtained in the fully
symmetric case (see Eq. (\ref{eq:f0_fn})) in the previous section if we set
$g_{2b}=g_b$, $g_{2s}=g_s$ and $\xi_2=\xi$ (the vdW contribution is the same in both cases).
This result follows straightforwardly  from the general quenched expressions (\ref{eq:force_quenched}) or (\ref{eq:f0_fn_app}) and again
reflects the fact that the quenched contribution is basically due to the interaction of  disorder charges
with their image charges (as these are the only  `charges' with which they are  `correlated' across the intervening gap).
This is why the disorder forces in the quenched case depend essentially  on the dielectric jump across
the bounding surfaces as discussed before.

\begin{figure}[t]
\begin{center}
\vspace{-2cm}
\includegraphics[angle=0,width=6.9cm]{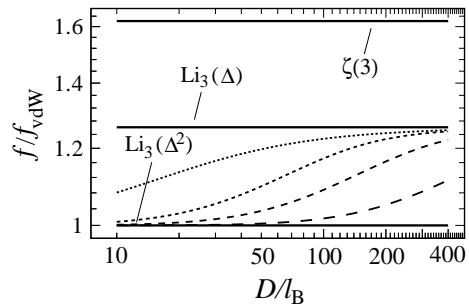}
\caption{Same
as Fig. \ref{fig:fig3}a but for two dissimilar dielectric slabs in
vacuum ($\varepsilon_m =1$) with one slab being {\em disorder
free} ($g_{1b} , g_{1s} =0$, $\xi_1 =0$) and the other slab
containing {\em annealed } charge disorder of variance $g_{2b} = 5
\times 10^{-8}\, {\mathrm{nm}}^{-3}$, $g_{2s} =0 $. Here we fix
$\varepsilon_p =10$ and vary the disorder correlation length as
$\xi_2 / l_{\mathrm{B}} = 0,200,10^3 , 10^4$ (from top
to bottom). The results in this case are bounded by the limiting
values given by Eqs. (\ref{eq:vdW}) and
(\ref{eq:f_star}) (solid lines labeled by
${\mathrm{Li}}_3(\Delta^2)$ and ${\mathrm{Li}}_3
(\Delta)$, respectively); see Eq. (\ref{eq:f_star_compare}). The top solid line
(labeled by $\zeta (3)$) is from Eq. (\ref{eq:metal}).}
\label{fig:fig6}
\end{center}
\vspace{-.7cm}
\end{figure}

\begin{figure*}[t]\begin{center}
	\begin{minipage}[b]{0.35\textwidth}\begin{center}
		\includegraphics[width=\textwidth]{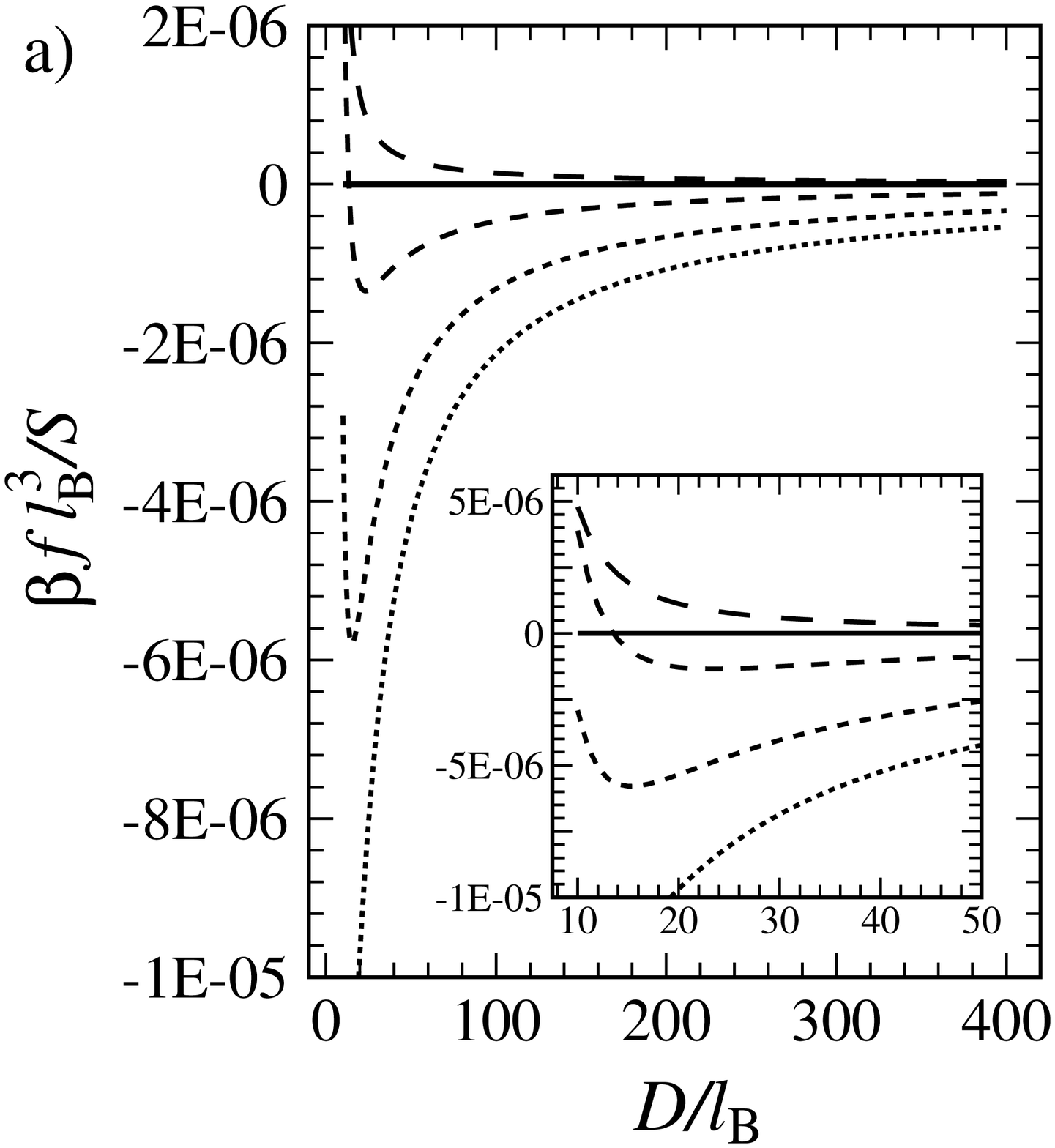} 
	\end{center}\end{minipage} \hskip-1cm
	\begin{minipage}[b]{0.35\textwidth}\begin{center}
		\includegraphics[width=\textwidth]{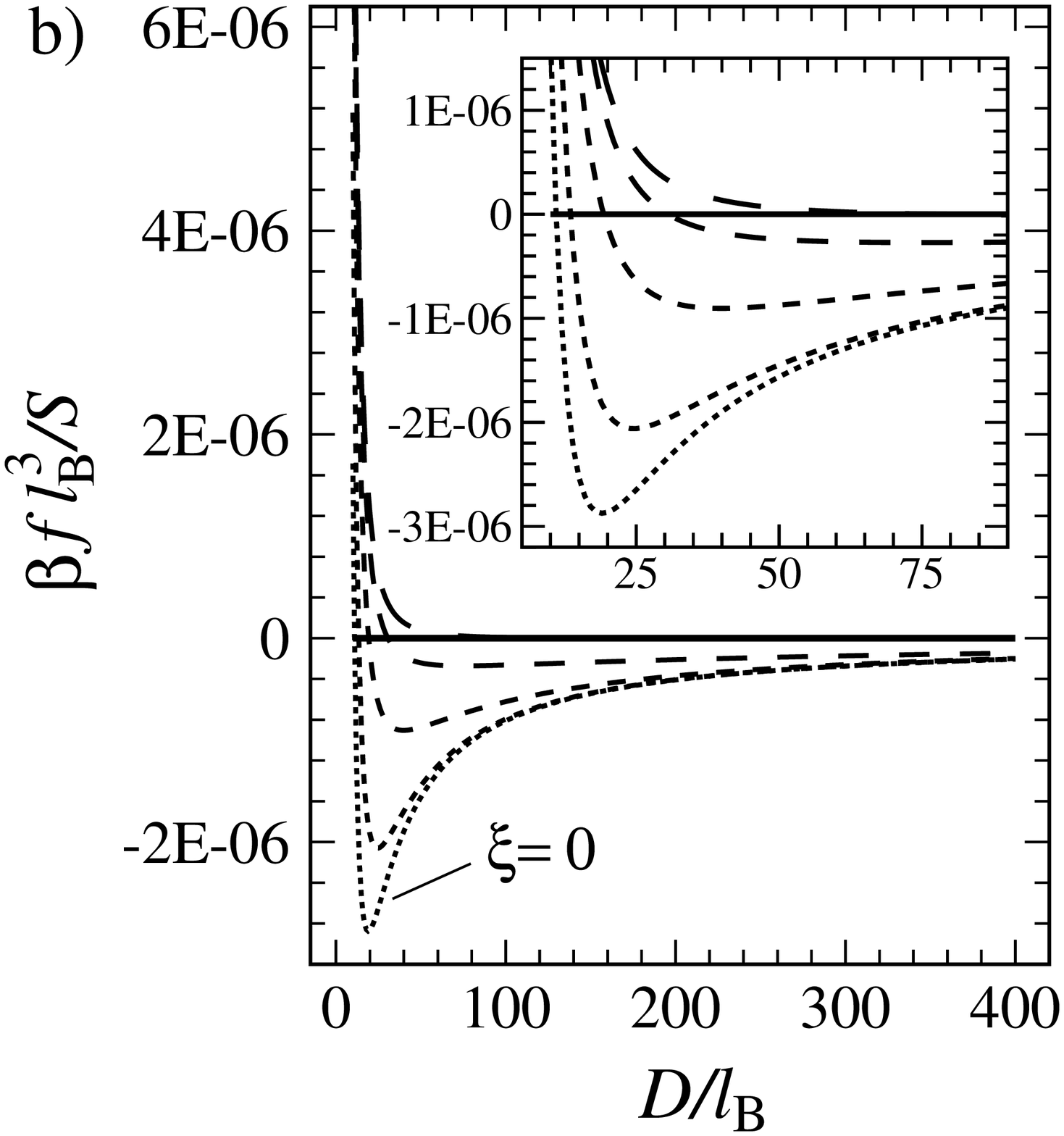} 
	\end{center}\end{minipage} \hskip-1cm
	\begin{minipage}[b]{0.35\textwidth}\begin{center}
		\includegraphics[width=\textwidth]{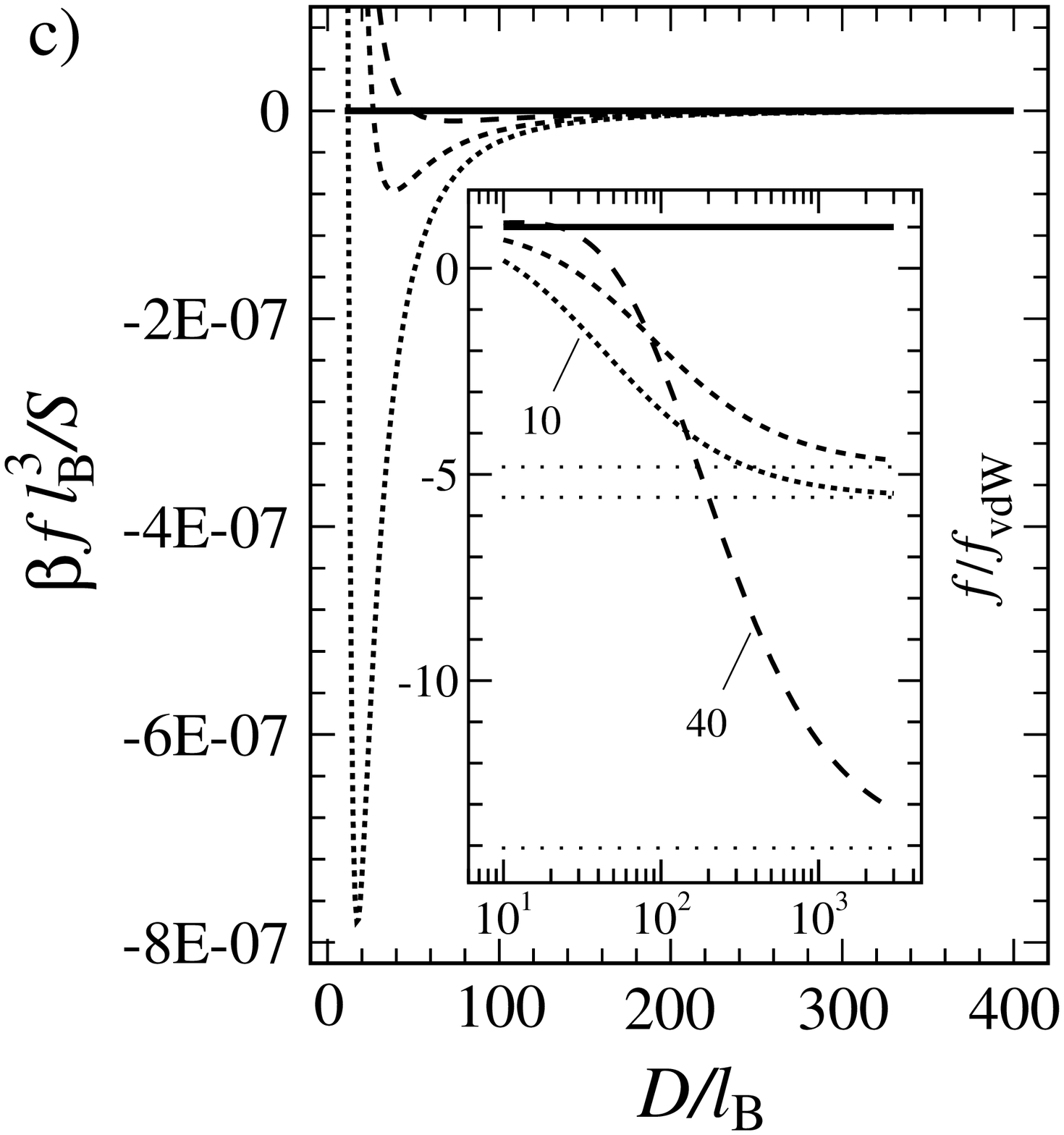}
	\end{center}\end{minipage} \hskip-1cm	
\caption{a) The rescaled total force, $\beta f l_{\mathrm{B}}^3
/S$ (Eq. (\ref{eq:force_quenched})), between two dissimilar
net-neutral dielectric slabs interacting across a medium of
dielectric constant $\varepsilon_m$ varying in the range
$\varepsilon_m = 10,15,25,40$ (dashed curves from bottom).
Here we have fixed the dielectric constant of the slabs as
$\varepsilon_1=5$, $\varepsilon_2=50 $, which contain uncorrelated
{\em quenched} disorder ($\xi_1 =\xi_2 =0$) of fixed disorder
variances $g_{1b}=g_{2b}= 5 \times 10^{-8} \,{\mathrm{nm}}^{-3}$ and
$g_{1s}=g_{2s}=0$. Inset shows a closer view of the region around
the minimum. b) Same as (a) but plotted for correlated quenched
disorder. Here we fix $\varepsilon_m =20$ and vary the correlation
length as $\xi / l_{\mathrm{B}} = 0, 20, 100,500, 10^4$,
which is taken to be equal in both slabs $\xi_1 =\xi_2 =\xi$. 
 Inset again shows a closer view of the region around
the minimum.  c) Same
as (a) but for (uncorrelated) {\em annealed} disorder obtained
from Eq. (\ref{eq:force_annealed}) for $\varepsilon_m = 10,20,40$. 
Inset shows the ratio of the
total force to the pure zero-frequency vdW force
(\ref{eq:vdW_full}) in the absence of charge disorder for a wider
range of separations. The horizontal dotted lines show
the limiting expression (\ref{eq:metal}).}
\label{fig:fig7}
\end{center}
\end{figure*}

In the annealed case, the disordered dielectric slab tends to
behave asymptotically as  a  perfect conductor, while the
disorder-free slab behaves as a dielectric material. This lowers
the {\em magnitude} of the maximum net force that can be achieved
in this system as compared with the case where both slabs bear
annealed charges; the latter case is obviously more favorable
thermodynamically as the system can achieve a lower free energy. 
The net annealed force in
vacuum still falls between two well-defined limits as shown in
Fig. \ref{fig:fig6}.
The two bounding limits are given by
\begin{equation}
  |f_{\mathrm{vdW}}| < |f_{\mathrm{annealed}}| <   |f_\ast|, \label{eq:f_star_compare}
\end{equation}
where $ f_{\mathrm{vdW}}$, Eq. (\ref{eq:vdW}), is obtained in the limit of weak disorder or at small separations, and $f_\ast$, defined as 
\begin{equation}
    \frac{\beta f_\ast}{S} = -\frac{{\mathrm{Li}}_3(\Delta)}{8\pi D^3},
        \label{eq:f_star}
\end{equation}
is obtained in the limit of strong disorder or large separation. Thus, in marked contrast with the fully symmetric case,  we find that
the large-distance interaction here becomes non-universal and could be either attractive ($\Delta>0$) or repulsive ($\Delta<0$).
This points to the possibility of {\em repulsive} total forces in the annealed case that will be discussed in the following section.

\begin{figure*}[t]\begin{center}
	\begin{minipage}[b]{0.35\textwidth}\begin{center}
		\includegraphics[width=\textwidth]{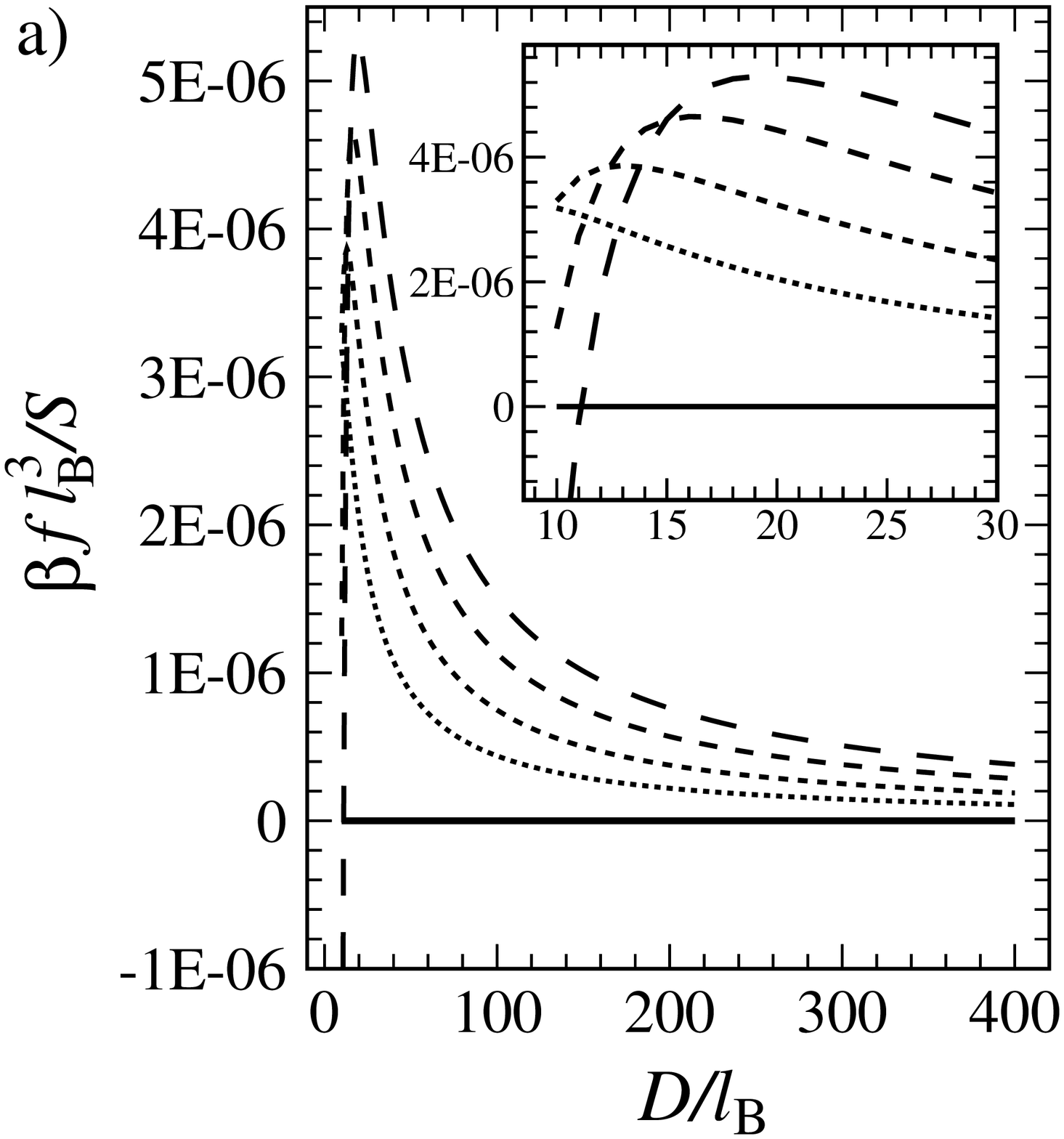} 
	\end{center}\end{minipage} \hskip-1cm
	\begin{minipage}[b]{0.35\textwidth}\begin{center}
		\includegraphics[width=\textwidth]{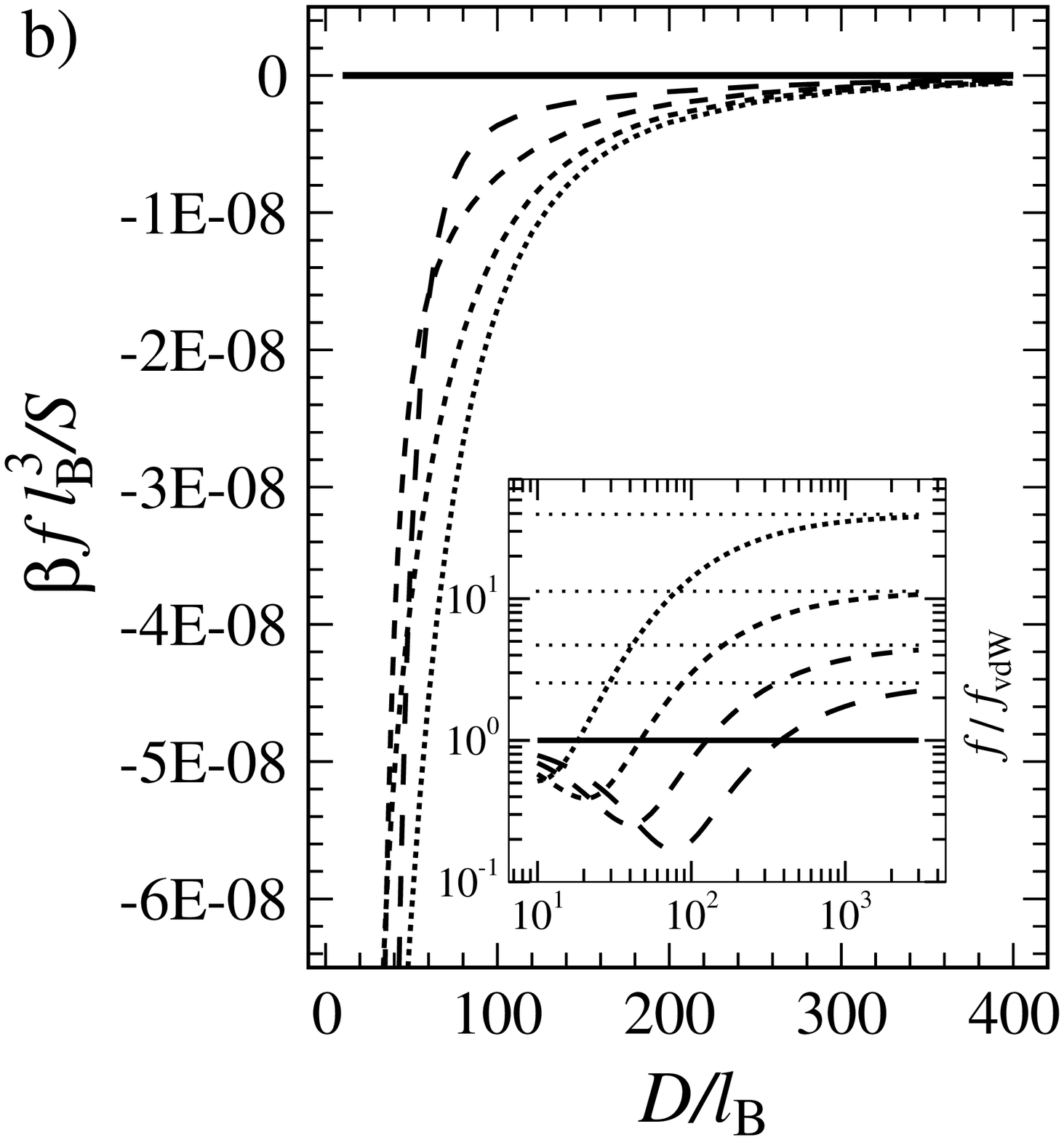} 
	\end{center}\end{minipage} \hskip-1cm
	\begin{minipage}[b]{0.35\textwidth}\begin{center}
		\includegraphics[width=\textwidth]{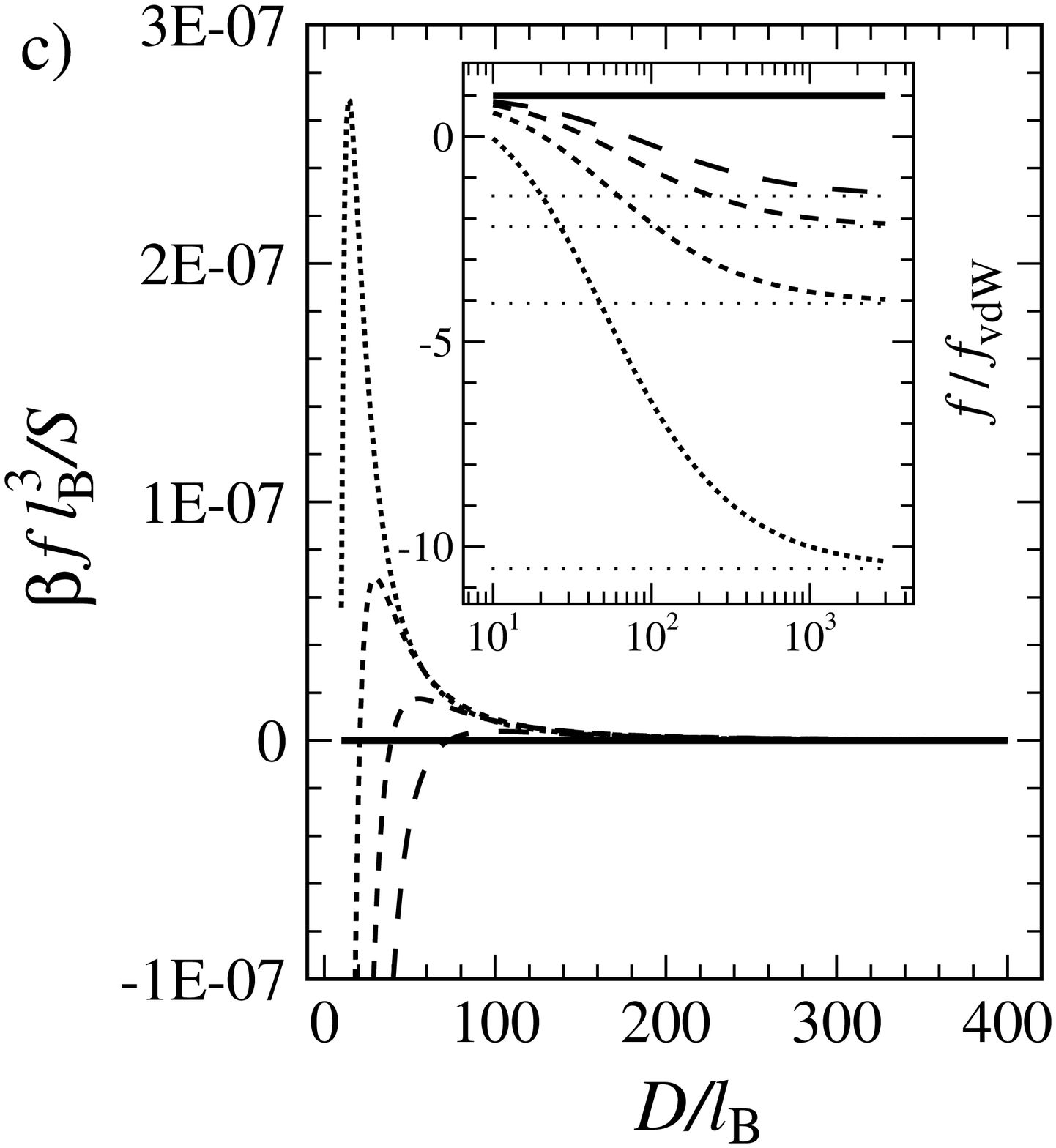}
	\end{center}\end{minipage} \hskip-1cm
\caption{a) The rescaled total force, $\beta f l_{\mathrm{B}}^3
/S$ (Eq. (\ref{eq:force_quenched})), between two dissimilar
net-neutral dielectric slabs interacting across a medium of
dielectric constant $\varepsilon_m$ varying in the range
$\varepsilon_m = 30,40,60,100$ (dashed curves from bottom).
Here we have fixed the dielectric constant of the slabs as
$\varepsilon_1=15$, $\varepsilon_2=25 $, which contain uncorrelated
{\em quenched} disorder ($\xi_1 =\xi_2 =0$) of fixed disorder
variances $g_{1b}=g_{2b}= 5 \times 10^{-8}\, {\mathrm{nm}}^{-3}$ and
$g_{1s}=g_{2s}=0$. Inset shows a closer view of the region around
the maximum. b) Same as (a) but for {\em annealed} disorder
obtained from Eq. (\ref{eq:force_annealed}). Inset shows the ratio
of the total force to the vdW force (\ref{eq:vdW_full}) for a
wider range of separations. c) Same as (a) but for one slab being
{\em disorder free} ($g_{1b}= g_{1s} =0$, $\xi_1 =0$) and the other
slab containing {\em annealed} charge disorder of variances $g_{2b} = 5
\times 10^{-8} \,{\mathrm{nm}}^{-3}$ and $g_{2s} =0$ (dashed curves
from top correspond to $\varepsilon_m = 30,40,60,100$). Inset
shows the ratio of the total force to the vdW force
(\ref{eq:vdW_full}) for a wider range of separations. }
\label{fig:fig8}
\end{center}
\end{figure*}

\subsection{Non-monotonic interaction between dissimilar slabs with $\varepsilon_1<\varepsilon_m<\varepsilon_2$}

It is clear that the  vdW force (\ref{eq:vdW_full}) becomes {\em repulsive} when the dielectric constant of the intervening medium is between that of the two semi-infinite slabs, i.e., $\varepsilon_1<\varepsilon_m<\varepsilon_2$ (such a case has been investigated experimentally in Ref. \cite{vdw_repulsive}). However, in this case, the force stemming from the quenched charge disorder can be {\em attractive} and thus, if present, may compete against
the repulsion due to the vdW forces. In order to investigate this situation further, let us first consider the case
of an uncorrelated disorder by setting the disorder correlation lengths equal to zero. Also without loss of generality,
we assume that the slabs contain only quenched bulk disorder
with equal variances $g_{1b}=g_{2b}=g_b$
in the two slabs. The net quenched force in this case reads 
\begin{equation}
\frac{ \beta {f}_{\mathrm{quenched}}}{S} =  -\frac{{\mathrm{Li}}_3(\Delta_1 \Delta_2)}{8\pi D^3}
 - \frac{g_{b} \ell_B \,\chi}{2(\varepsilon_1 + \varepsilon_2) D},
 \label{eq:quenched_1m2}
\end{equation}
where we have defined
\begin{equation}
\chi =\frac{(\varepsilon_1 - \varepsilon_m) }{ \varepsilon_2 + \varepsilon_m}
 + \frac{(\varepsilon_2 - \varepsilon_m) }{ \varepsilon_1 + \varepsilon_m}.
\end{equation}
Obviously, the contribution from bulk disorder (second term in Eq. (\ref{eq:quenched_1m2})) can change sign as the
dielectric constant $ \varepsilon_m$ is varied in the range $\varepsilon_1<\varepsilon_m<\varepsilon_2$, while
the vdW term (first term) remains always repulsive. It is easy to see that for $\varepsilon_m\rightarrow \varepsilon_1$,
we have $\chi>0$ and thus an attractive disorder force, and   for $\varepsilon_m\rightarrow \varepsilon_2$,
we have $\chi<0$ and thus a repulsive disorder force.
This behavior is shown in Fig. \ref{fig:fig7}a, where
$\varepsilon_1=5$, $\varepsilon_2=50$ are fixed and
$\varepsilon_m$ varies in the range $\varepsilon_m = 10, 15,
25, 40$ (dashed curves from bottom). In accordance with our
findings in the symmetric case, the large distance behavior is
always dominated by the disorder contribution (as the disorder
force decays more weakly with the separation), while the repulsive vdW force
in this case plays the role of a stabilizing force at small
separations.

The resulting effect is that the total force varies {\em non-monotonically} and vanishes at a finite distance, $D_0$, between the two slabs
given in the present case by
\begin{equation}
 D_0^2 = -\frac{{\mathrm{Li}}_3(\Delta_1 \Delta_2)}{4\pi g_{b} \ell_B \,\chi}(\varepsilon_1 + \varepsilon_2).
 \label{eq:D0}
 \end{equation}
This represents a stable  `equilibrium' separation (bound state)
between the two slabs corresponding to a minimum in the
interaction free energy. On the other hand, the maximum attractive
force due to the influence of quenched disorder is reached at a
larger separation,
\begin{equation}
D_{\mathrm{max}}=\sqrt{3}D_0.
 \label{eq:Dmax}
\end{equation}
These results may be used to optimize the thickness of the intervening medium in order to achieve the maximum or minimum force
magnitude between the slabs.

In the presence of a correlated disorder, the disorder-induced
effects weaken (Section \ref{sec:symm_quenched}) and thus, as the
disorder correlation is increased, the attractive tail as well as
the stable bound state are gradually washed out as shown in Fig.
\ref{fig:fig7}b.

We find similar features when the disorder charges are annealed as
seen in Fig. \ref{fig:fig7}c. However, in line with our finding in
the symmetric case, the net force falls off more rapidly with the
separation when the disorder charges are annealed. The large
distance behavior coincides again with the universal expression
(\ref{eq:metal}) and can be much larger in magnitude than the pure
vdW force (\ref{eq:vdW_full}) when $\varepsilon_m$ tends to the
larger dielectric constant $\varepsilon_2$ as shown in the inset
of Fig. \ref{fig:fig7}c.

\subsection{Non-monotonic interaction between dissimilar slabs with $\varepsilon_1, \varepsilon_2<\varepsilon_m$}

In the case where the intervening medium has a higher dielectric
constant than the two slabs, $\varepsilon_1,
\varepsilon_2<\varepsilon_m$, the vdW effect leads to an {\em
attractive} force between the two slabs, Eq. (\ref{eq:vdW_full}),
which again dominates at small separations. If the  disorder is
quenched, the force generated by the disorder can become {\em
repulsive} and thus lead to a potential barrier and a long
repulsive tail at large separations. This is shown in Fig.
\ref{fig:fig8}a, for the case of  two slabs with uncorrelated bulk
disorder of  equal variances ($g_{1b}=g_{2b}=g_b$),
 where we have fixed $\varepsilon_1=15$, $\varepsilon_2=25$  and vary $\varepsilon_m$ in the range
$\varepsilon_m = 30, 40, 60, 100$ (dashed curves from bottom).
The total force in this case is given again by Eq.
(\ref{eq:quenched_1m2}). Note that the disorder-induced force
(second term) is always repulsive in the case with
$\varepsilon_1, \varepsilon_2<\varepsilon_m$ (i.e., $\chi<0$).
Thus, the separation distance $D_0$, Eq. (\ref{eq:D0}),
corresponds to an unstable  `equilibrium' distance between the two
slabs and $D_{\mathrm{max}}$ gives the distance at which the
maximum repulsive force due to the influence of quenched disorder
is achieved.

The above features change dramatically if the disorder charges are
assumed to be annealed. In fact, the net annealed force appears to
follow a trend similar to what one expects from the pure vdW
force, i.e., in contrast with the quenched case, the net annealed
force turns out to be attractive and vary monotonically with the
separation (Fig. \ref{fig:fig8}b). However, the ratio of the net
force to the vdW force (\ref{eq:vdW_full}) shows that the relative
magnitude of the force can vary non-monotonically and deviate
significantly from the underlying vdW force  (Fig. \ref{fig:fig8}b, inset). 
In particular, one
observes that the relative net force may be enhanced here by an
order of magnitude at large separations. This is again due to
fluctuations of annealed charges that can redistribute in the
slabs in such a way as to minimize the free energy and thus favor
a higher attractive force. This effect appears to be much stronger
in an asymmetric system than in a symmetric system (Figs.
\ref{fig:fig3} and \ref{fig:fig6}), where the net force remains of
the same order as the pure vdW force.

In the preceding discussion, we assumed that the two slabs have
similar annealed disorder variances, which leads to a purely
attractive force between the slabs for $\varepsilon_1,
\varepsilon_2 ~ < ~ \varepsilon_m$. It turns out that the annealed
disorder can also lead to a {\em repulsive} force in this case
provided that the charge is distributed asymmetrically between the
two slabs. In Fig. \ref{fig:fig8}c, we show the results for the
case where one of the slabs is disorder free, but the other slab
contains annealed charge disorder of finite variance. As seen, one
can achieve an interaction (potential) barrier even with the
annealed charges. This effect is stronger when $\varepsilon_m$ is
smallest and disappears when $\varepsilon_m \rightarrow \infty$.

\section{Discussion}
\label{sec:discussion}

In this paper, we have shown that the effective interaction induced by
the quenched or annealed monopolar charge disorder can give rise to various novel
features in the overall interaction between two net-neutral semi-infinite dielectric slabs depending on the detailed assumptions about the disorder and the dielectric inhomogeneities in the system. The quenched case and the annealed case of disorder differ in the sense that in the former the disordered charges
are  frozen and can not fluctuate, while in the latter the disordered charges are subject to thermal fluctuations and adapt themselves to  minimize the  free energy of the system. 
Our analysis is based on recent
developments which unify the salient features of the zero-frequency vdW interaction and
the physics of the charge disorder \cite{ali-rudi,mama,PRL}. Based on these developments 
we argue that, in the case of two dissimilar slabs, the charge disorder-induced electrostatic interaction can have an opposite sign to the zero-frequecny
vdW interaction and can thus give rise 
to a {\em non-monotonic} net interaction between  the slabs. 
The most surprising features of our analysis pertain to the case of 
disorder-induced
interactions across a medium of higher dielectric constant than that of the two slabs. In this case the 
net force may become strongly repulsive and can lead to a
potential barrier for stiction when combined with the attractive vdW force.
In the opposite case, where the intervening medium  has a
dielectric constant in between that of the two slabs, the 
disorder may generate a long-range attraction, which opposes
the {\em repulsive} vdW force and can thus promote a stable bound state for the two bounding 
dielectrics. These salient features of the disorder-induced interactions are due to the slower 
decay of these interactions and their stronger dependence on
the dielectric inhomogeneities in the system as 
compared to the corresponding vdW interaction. 



This comparison between the disorder-induced and the classical zero-frequency 
vdW interaction is in order since both of them are expected to be valid in the regime of large 
intersurface separations  (or high temperatures)
 as considered in this work and which is also 
 most  relevant to recent experiments \cite{kim}. 
The precise correction presented by the higher-order Matsubara frequencies 
\cite{parsegianbook} is highly material specific, but its magnitude (relative to the zero-frequency term) is typically small for the most part of the separation range considered here and remains negligible in comparison with the  disorder effects. 


Being disorder induced one should be in principle also able to compute the corresponding statistical moments of the disorder interactions 
averaged over the disorder. Note that these fluctuations over the disorder will be different from 
the thermal fluctuations of the Casimir force studied in Ref. \cite{golest} which are present even in the absence of disorder.

\section{Acknowledgments}

D.S.D. acknowledges support from the Institut Universitaire de France. R.P. acknowledges support from ARRS through the program P1-0055 and the research project J1-0908. A.N. is supported by a Newton International Fellowship from the Royal Society, the Royal Academy of Engineering, 
and the British Academy. 
J.S. acknowledges generous support by J. Stefan Institute (Ljubljana) 
provided for a visit to the Institute. 

\appendix

\section{General expression for the force in the quenched case}

In the general case where the two slabs have different dielectric constants and charge disorder parameters, we can write the
total force, Eq. (\ref{eq:force_quenched}),  as a series expansion in powers of $\xi_{i\alpha}/D$ (where for the two slabs, $i=1,2$,
and bulk and surface disorder, $\alpha=b, s$), i.e.
\begin{equation}
f_{\mathrm{quenched}} =  f^{(0)} + \sum_{n=1}^{\infty} \big(f_1^{(n)} + f_2^{(n)}\big),
\label{eq:f0_fn_app}
\end{equation}
where we obtain
\begin{eqnarray}
 \frac{ \beta {f}^{(0)}}{S}&\! = \!&
  - \frac{g_{1b} \ell_B (\varepsilon_2 + \varepsilon_m)\Delta_2 }{2 (\varepsilon_1 + \varepsilon_m)(\varepsilon_1 + \varepsilon_2) D}
  - \frac{ g_{1s} \ell_B \varepsilon_m \log (|1-\Delta_1 \Delta_2|) }{(\varepsilon_1+\varepsilon_m)^2 \Delta_1 D^2}  \nonumber\\
&&  \hspace{-1.4cm}
 - \frac{g_{2b} \ell_B (\varepsilon_1 + \varepsilon_m) \Delta_1}{2 (\varepsilon_2 + \varepsilon_m)(\varepsilon_1 + \varepsilon_2) D}
- \frac{  g_{2s} \ell_B \varepsilon_m \log (|1-\Delta_1 \Delta_2|) }{(\varepsilon_2+\varepsilon_m)^2 \Delta_2 D^2},
\end{eqnarray}
and
\begin{eqnarray}
 \frac{ \beta {f}_1^{(n)}}{S} &\!= \!&
  - \frac{2 g_{1b} \ell_B \varepsilon_m}{(\varepsilon_1+\varepsilon_m)^2 \Delta_1 D}
\sum_{n=1}^{\infty} (-1)^n \frac{\xi_1^{2n}}{D^{2n}} C_{2 n +1} \mathrm{Li}_{2 n} (\Delta_1 \Delta_2)\nonumber\\
&& \hspace{-1.4cm}
- \frac{4 g_{1s} \ell_B \varepsilon_m}{(\varepsilon_1  +  \varepsilon_m)^2  \Delta_1
D^2}  \sum_{n=1}^{\infty} (-1)^n \frac{\xi_1^{2n}}{D^{2n}} C_{2 n +2} \mathrm{Li}_{2 n+1}(\Delta_1 \Delta_2).
\end{eqnarray}
The expression for $ f_2^{(n)}$ is obtained simply by replacing the subindex 1 with 2 and vice versa.


\end{document}